\documentclass[twocolumn,aps,amssymb,footinbib]{revtex4-1}
\usepackage{amssymb}
\usepackage{graphicx}
\usepackage{amsmath}
\usepackage{times}
\usepackage{color}
\usepackage{subfigure}
\usepackage{bm}

\newcommand{\xx}{\mathbf{x}}
\newcommand{\rr}{\mathbf{r}}
\newcommand{\qq}{\mathbf{q}}
\newcommand{\kk}{\mathbf{k}}

\newcommand{\GG}{\mathcal{G}}

\newcommand{\aprp}{a_{\perp}}
\newcommand{\RPA}{RPA$_\mathrm{ns}$~}

\begin{document}

\begin{abstract}
We study the in-plane and out-of-plane density ordering instabilities of quasi-two-dimensional fermionic polar molecules in single-layer and multi-layer configurations. We locate the soft modes by evaluating linear response functions within the conserving time-dependent Hartree-Fock (TDHF). The short-range exchange effects are taken into account by solving the Bethe-Salpeter integral equation numerically. An instability phase diagram is calculated for both single-layer and multi-layer systems and the unstable wave-vector is indicated. In all cases, the in-plane density wave instability is found to precede the out-of-plane instability. The unstable wave-vector is found to be approximately twice the Fermi wave-vector of one of the subbands at a time and can change discontinuously as a function of density and dipolar interaction strength. In multi-layer configurations, we find a large enhancement of density wave instability driven by dilute quasiparticles in the first excited subband. Finally, we provide a simple qualitative description of the phase diagrams using a RPA-like approach. Compared to previous works done within the RPA approximation, we find that inclusion of exchange interactions stabilize the normal liquid phase further and increase the critical dipolar interaction strength corresponding to the onset of density-wave instability by over a factor of two.
\end{abstract}

\title{Density ordering instabilities of quasi-two-dimensional fermionic polar molecules in single-layer and multi-layer configurations: exact treatment of exchange interactions}

\author{Mehrtash Babadi$^1$, Eugene Demler$^1$}
\affiliation{
$^1$ Physics Department, Harvard University, Cambridge, Massachusetts 02138, USA
}
\maketitle

\section{Introduction} 
The field of ultracold atoms has witnessed a rapid progress in the past decade. Much of this experimental and theoretical progress has been motivated by the prospect of realizing novel strongly correlated quantum phases and exploring the consequences of strong interactions~\cite{Lewenstein2007,Bloch2008,Ketterle2008}. One of the latest breakthroughs in this direction is the experimental realization of nearly quantum degenerate gases of fermionic polar molecules. By association of atoms via a Feshbach resonance to form deeply bound ultracold molecules~\cite{Lang2008,Deiglmayr2008}, a nearly degenerate gas of KRb polar molecules has been recently realized in their rotational and vibrational ground state~\cite{Ospelkaus2008,Ni2008,Ni2009,Ospelkaus2010,Ni2010}. The molecules can be polarized by the application of a dc electric field, resulting in strong dipole-dipole inter-molecular interactions.

At the time this paper is being written, the coldest realized gas of polar fermionic molecules has a temperature of $1.4\,T_F$ in the experiments of the group at JILA~\cite{Ospelkaus2008,Ni2008,Ni2009,Ospelkaus2010,Ni2010}, where $T_F$ is the Fermi temperature. A major obstacle toward further evaporative cooling of a large class of bi-alkali polar molecules is the existence of an energetically allowed two-body chemical reaction channel~\cite{Zuchowski2010}, resulting in significant molecule losses in two-body scatterings. In a low temperature gas composed of a single hyperfine state, Fermi statistics blocks scatterings in the {\it s}-wave channel and the majority of scatterings take place through the {\it p}-wave channel. In a three-dimensional gas, the attractive head-to-tail dipolar interactions soften the {\it p}-wave centrifugal barrier and increase the cross section of reactive collisions. The rate of chemical reactions can be effectively suppressed by loading the gas into a one-dimensional optical lattice (or trap) and aligning the dipoles perpendicular to the formed quasi-two-dimensional layers, also known as pancakes. In such geometries, the incidence of head-to-tail scatterings is effectively suppressed due to the transverse confinement of the gas on one hand, and reinforcement of the {\it p}-wave barrier due to repulsive side-by-side dipolar interactions on the other hand~\cite{Ospelkaus2010,Quemener2010,Quemener2011}. Therefore, the preferred geometry to study reactive polar molecules is in tightly confined two-dimensional layers.

\begin{figure}[t!]
\center
\includegraphics[scale=0.75]{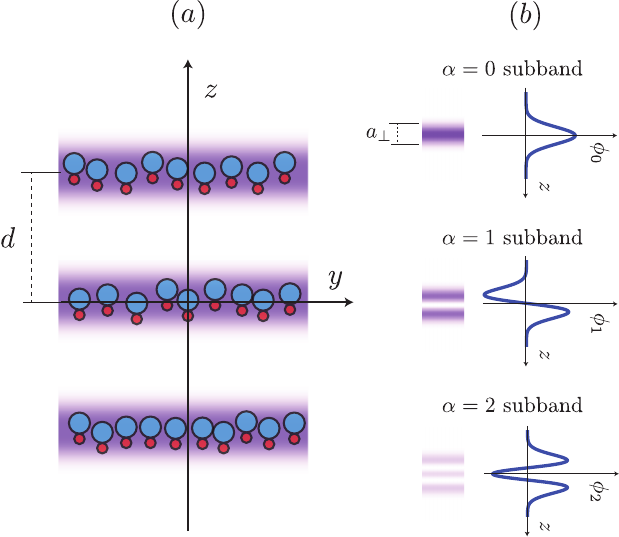}
\caption{(Color online) (a) A multi-layer system of quasi-two-dimensional polar molecules. The dipoles are aligned perpendicular to the $x$-$y$ plane by the application of a strong dc electric field. The inter-layer spacing is $d$. The $x$-axis is perpendicular to the plane of the plot. (b) each quasi-two-dimensional layer is composed of multiple subbands, corresponding to the transverse wavefunctions of the trap. For symmetric lattice potentials, the subbands have a well-defined parity with respect to the reflection about the $x$-$y$ plane. The trap confinement width, $a_\perp$, is shown in the figure.}
\label{fig:intuition2}
\end{figure}

In such geometries, the energy levels of particles is quantized due to the transverse confining potential and each quasi-two-dimensional layer can be thought of as a collection of two-dimensional energy subbands (see Fig.~\ref{fig:intuition2}). Since higher subbands have a larger transverse spreading, it is expected that occupation of higher subbands will increase the rate of head-to-tail collisions and consequently, the molecule loss rate. However, it has been recently shown that the two-body chemical reactions will still be significantly suppressed even if the first few subbands are filled, due to Pauli blocking~\cite{Quemener2011}. The occupation of higher subbands does not impose any difficulty on experiments with non-reactive species such as $\mathrm{NaK}$, $\mathrm{NaRb}$, $\mathrm{NaCs}$, $\mathrm{KCs}$, and $\mathrm{RbCs}$~\cite{Zuchowski2010}. The possibility of going beyond the single-subband limit opens a new window toward experimental and theoretical exploration of many-body physics of quasi-two-dimensional fermionic systems with anisotropic interactions.\\

In contrast to the isotropic short-range interactions in ultracold atomic gases realized using a {\it s}-wave Feshbach resonance~\cite{Lewenstein2007,Bloch2008,Ketterle2008}, the long-range and anisotropic nature of electric dipole-dipole interactions in ultracold gases of polar molecules allows the experimental realization of a wider range of physical phenomena. In particular, the repulsive side-by-side dipole-dipole interactions in layered stacks of polar molecules can lead to spontaneous translational symmetry breaking and formation of density ordered phases for strong interactions. We define the ratio of the typical interaction over the kinetic energy, $r_d$, as a dimensionless measure of the strength of dipolar interaction:
\begin{equation}\label{eq:rd}
r_d \equiv \frac{m\,D^2\,n^{1/2}}{\hbar^2},
\end{equation}
where $D$ is the electric dipole moment of a single molecule, $m$ is the molecular mass and $n$ is the two-dimensional density. It is noticed that in contrast to the electron gas, the interaction energy is dominants at higher densities for fixed dipole moments and as a result, the density ordered phases are expected to appear at higher densities.\\

Recently, Yamaguchi {\it et al.}~\cite{Yamaguchi2010} and Sun {\it et al.}~\cite{Sun2010} have independently studied the density-wave (DW) instability in a strictly two-dimensional layer of polar molecules in the random phase approximation (RPA). At zero temperature and zero tilt angle of dipoles with respect to the confining 2D plane, their results indicate that the DW instability occurs for $r_d \approx 0.17$. The former study treats the self-energy corrections within an approximate variational method and first-order perturbation theory~\cite{Yamaguchi2010}. The second study neglects the self-energy corrections altogether, however, present a rigorous proof for the necessity of DW instability for strong enough interactions and predict the nature of the density ordered phases at different tilt angles~\cite{Sun2010}. Both studies neglect the exchange interaction effects beyond the cancellation of the {\it s}-wave component of dipolar interaction in their calculations. Since the interactions need to be appreciably strong for the density ordered instabilities to occur and that the ordering wave-vector is in order of the Fermi wave-vector, we expect that inclusion of short-range exchange effects will result in a significant quantitative correction to the results of the cited works.\\

The simplest self-consistent and conserving many-body approximation that respects the Fermi statistics is the time-dependent Hartree-Fock approximation (TDHF)~\cite{Baym1961}, also known as the generalized random phase approximation (GRPA)~\cite{NozieresPines}. We have recently studied the band renormalization and collective modes of quasi-two-dimensional polar molecules within the TDHF approximation~\cite{Babadi2011}. In this paper, we study the density ordering instabilities of the liquid phase of quasi-two-dimensional polar molecules in single-layer and multi-layer configurations. Throughout this study, we assume that the layers are well-separated such that the inter-layer tunneling can be neglected.\\

\begin{figure}[t!]
\center
\includegraphics[scale=0.7]{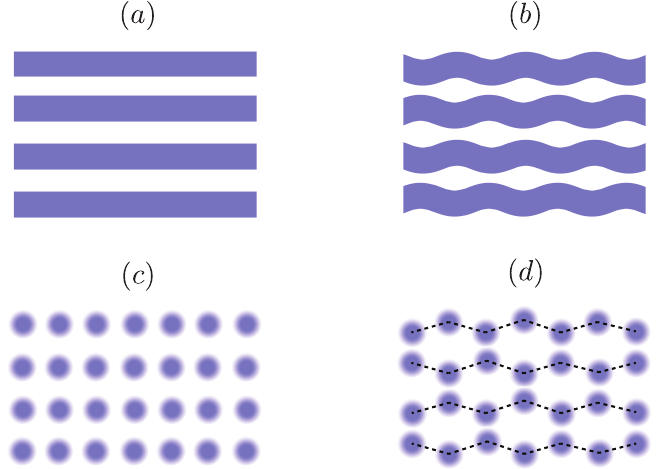}
\caption{(Color online) A schematic representation of the homogeneous liquid and density ordered phases of a multi-layer system of quasi-two-dimensional polar molecules. (a) the liquid phase is characterized by uniform in-plane and out-of-plane density in each layer. (b) the ripplon phase is characterized by out-of-plane density modulations and uniform in-plane projected density. The $\mathbb{Z}_2$ reflection symmetry may be broken (as shown here) if the mixing occurs between subbands of even and odd parity. In that case, the energetically favorable configuration corresponds to a $\pi$ phase shift between even and odd layers due to the inter-layer attraction. (c) the density-wave phase is characterized by broken in-plane translational symmetry. Wigner crystals, striped and bubble phases are examples of density-wave phases. (d) the zigzag crystal phase is characterized by broken in-plane translation symmetry and presence of out-of-plane density modulations.}
\label{fig:phases}
\end{figure}

We consider instability toward two types of density ordered phases: in-plane density-wave phase and the {\em ripplon} phase (see Fig.~\ref{fig:phases}). The in-plane density-wave phase is characterized by broken in-plane translational symmetry~(see Fig.~\ref{fig:phases}c). The ripplon phase is a reminiscent of the spin density-wave (SDW) phase of electronic systems~\cite{Overhauser1962}, where quasiparticles of different subbands play the role of different spin states. Since quasiparticles in difference subbands have different transverse wavefunctions, their mixing results in out-of-plane density modulations (see Fig.~\ref{fig:phases}b). In the ripplon phase, the $\mathrm{Z}_2$ reflection symmetry about the confining plane may be broken if the mixing occurs between subbands of even and odd parity. When the translational symmetry of the in-plane projected density is broken in addition to the presence of subband mixing, we denote the phase by {\em zigzag} phase (see Fig.~\ref{fig:phases}d). Quantum zigzag transition in one-dimensional ion chains has been a subject of active theoretical and experimental investigations in the past few years~\cite{Meyer2009,Fishman2008,Astrakharchik2008,Shimshoni2011,Meng2011}.\\

We adhere to the mode-softening paradigm of phase transition and look for the instabilities by monitoring various density-density response functions in the normal phase as the interaction and trap strengths are varied. The softening of a density-wave mode results in development of a sharp peak in its corresponding response function, which eventually promotes to a singularity when the liquid phase becomes unstable against density fluctuations. 

We remark that the TDHF approximation adopted in this study is essentially an analysis of Gaussian fluctuations about the mean-field liquid phase. In many cases, analyses at the mean-field level underestimate the stability of symmetric phases and predict transition to the symmetry broken phases too early. Inclusion of correlation effects usually enhance the stability of the unbroken phase beyond mean-field predictions. Thus, the instability criterion resulting from TDHF response functions is most likely only a signal for the formation of short-range correlations in the liquid phase, with the phase transition to the density ordered phase occurring at stronger interactions. Nevertheless, the mean-field stability analysis is a first step and an indispensable guide in constructing more elaborate approximations. We discuss this issue with more detail in Sec.~\ref{sec:disc}.\\

This paper is organized as follows: we review the microscopic model for single-layer and multi-layer configurations in Sec.~\ref{sec:model}. The TDHF formalism for multi-subband and multi-layer systems is described in Sec.~\ref{sec:response} and the instability phase diagrams are presented in Sec.~\ref{sec:res}. An approximate RPA-like theory is developed in Sec.~\ref{sec:RPA} using which the qualitative features of the obtained instability phase diagrams are explained. Finally, we discuss the results in Sec.~\ref{sec:disc} and compare our results with the previous works. The derivation of analytical expressions for the effective inter-layer and inter-subband dipolar interactions and the details of our numerical method are provided in the appendices.

\section{The Microscopic Model}\label{sec:model}
We start our analysis by reviewing the microscopic model describing fermionic polar molecules of mass $m$, prepared in a single hyperfine state, and loaded in a one-dimensional optical lattice. For concreteness, we assume that the optical lattice is along the {\it z}-axis. Also, we assume throughout this paper that all of the dipoles are aligned perpendicular to the trap plane by the application of a strong external d.c. electric field (see Fig.~\ref{fig:intuition2}). We work within the units $\hbar = 2m = 1$ unless these quantities appear explicitly. We also denote 3D and in-plane 2D coordinates by $\rr$ and $\xx$ respectively.

The second-quantized Hamiltonian describing the system can be written as:
\begin{eqnarray}\label{eq:H}
\mathcal{H} &=& \int \mathrm{d}^2\rr\,\psi^{\dagger}(\rr)\left[-\nabla^2 + V_{\mathrm{lat.}}(z)\right]\,\psi(\rr)\nonumber\\
&&+\int\int \mathrm{d}^2\rr\,\mathrm{d}^2\rr'\,\psi^{\dagger}(\rr) \psi^{\dagger}(\rr') V_{\mathrm{dip.}}(\rr-\rr') \psi(\rr') \psi(\rr),\nonumber\\
\end{eqnarray}
where $\psi(\rr)$ is the fermion annihilation operator and $V_{\mathrm{lat.}}(z)$ and $V_{\mathrm{dip.}}(\rr-\rr')$ denote the optical potential and dipolar interaction respectively:
\begin{align}\label{eq:optpot}
V_{\mathrm{lat.}}(z) =& V_0\,\sin^2(2\pi z/\lambda),\\
V_{\mathrm{dip.}}(\rr) =& \frac{D^2}{|\rr|^5}\,\left(|\rr|^2 - 3z^2\right),
\end{align}
where $\lambda$ is the wavelength of the optical lattice lasers. The gas in the optical lattice can be thought of as a stack of quasi-two-dimensional layers separated by a distance $d=\lambda/2$. In practice, there are a finite number of layers present in the sample. We denote the number of layers by $N_l$ and assume a periodic boundary condition along $z$ direction in order to eliminate the surface effects. We also denote the transverse size of the stack by $L \equiv N_l d$. 

The fermion operator can be conveniently expanded in Wannier's basis in $z$ direction and plane-wave basis in $x$-$y$ plane:
\begin{equation}\label{eq:fermop}
\psi(\rr) = \sum_{\kk} \sum_{\alpha=1}^{\infty} \sum_{n=1}^{N_l} w_{\alpha n}(z)\,\frac{e^{i\kk\cdot\xx}}{\sqrt{A}}\,c_{n \alpha, \kk}
\end{equation}
where $A$ is the area of system in $x-y$ planes, $w_{\alpha n}(z)$ denotes the Wannier's wavefunction of the band $\alpha$, with its center shifted to $n$'th well of the optical lattice and $c_{n \alpha, \kk}$ annihilates a fermion in layer $n$, subband $\alpha$ and with momentum $\kk$. We omit the limits in the summations over layer and subband indices in the rest of the paper for brevity. Plugging the expansion of the fermion operator into Eq.~(\ref{eq:H}), we get:
\begin{eqnarray}\label{eq:Hexp}
\mathcal{H} &=& \sum_{\kk}\sum_{mn}\sum_{\alpha\beta} \left(|\kk|^2 + J_{\alpha\beta}^{mn}\right)\,c^{\dagger}_{m\alpha,\kk} c^{\phantom{\dagger}}_{n\beta,\kk} + \frac{1}{2A}\,\sum_{\kk_1,\kk_2,\qq}\nonumber\\
&& \sum_{\alpha\beta;\gamma\lambda}\sum_{mn;rs}
\mathcal{V}_{\alpha\beta;\gamma\lambda}^{mn;rs}(\qq)\,c^{\dagger}_{m\alpha,\kk_1+\qq} \, c^{\dagger}_{r\gamma,\kk_2-\qq} c^{\phantom{\dagger}}_{s\lambda,\kk_2} \, c^{\phantom{\dagger}}_{n\beta,\kk_1},\nonumber\\
\end{eqnarray}
where:
\begin{equation}\label{eq:J}
J_{\alpha\beta}^{mn} = \int\mathrm{d}z\,w^*_{\alpha n}(z)\left[-\frac{\mathrm{d}^2}{\mathrm{d}z^2} + V_{\mathrm{lat.}}(z)\right]\,w_{\beta n}(z),
\end{equation}
and:
\begin{eqnarray}\label{eq:V}
\mathcal{V}_{\alpha\beta;\gamma\lambda}^{mn;rs}(\qq)&=&\int \mathrm{d}^2\xx\,e^{-i\qq\cdot(\xx-\xx')}\int \int \mathrm{d}z \, \mathrm{d}z' \, w_{\alpha m}^*(z)\nonumber\\&&\times\,w_{r\gamma}^*(z')\,w_{s\lambda}(z')\,w_{n\beta}(z)\,V_{\mathrm{dip}}(\rr-\rr').\nonumber\\
\end{eqnarray}

In order to simplify the analysis, we assume that the optical potential is deep and that its minima are well separated, so that we can neglect inter-layer tunneling effects.  We call this limit the independent layers (IL) limit. The subsequent results presented in this paper are all within this limit. In the IL limit, the optical potential (Eq.~\ref{eq:optpot}) can be expanded to quadratic order about its minima:
\begin{equation}
V_{\mathrm{lat.}} \simeq \sum_{n=0}^{N_l-1} \frac{1}{2}\,m\,\omega_\mathrm{trap}^2(z-nd)^2,
\end{equation}
where $\omega_\mathrm{trap} = (2\pi/\lambda)\sqrt{2V_0/m}$ in the effective trap frequency of each well. The Wannier's wavefunctions can also be approximated by shifted harmonic oscillator wavefunctions:
\begin{eqnarray}\label{eq:HO}
w_{\alpha n}(z) &\simeq& \phi_{\alpha}(z-nd),\nonumber\\
\phi_{\alpha}(z) &=& \frac{e^{-z^2/2 a_{\bot}}}{\sqrt{\pi^{1/2}\,\alpha!\,2^\alpha\,a_\bot}}\,H_{\alpha}\left(z/a_{\bot}\right),
\end{eqnarray}
where $H_\alpha$ is the Hermite polynomial of degree $\alpha$, and $a_\bot$ is the transverse spreading of the lowest subband which is related to the parameters of the optical lattice as:
\begin{equation}\label{eq:aperp}
a_{\perp}^2 = \frac{\hbar}{m\omega_{\mathrm{trap}}} = \frac{h\lambda}{\sqrt{2mV_0}}.
\end{equation}
The conditions of the IL limit is met if the transverse spreading of the Wannier's wavefunctions is smaller than the inter-layer separation, i.e. $a_\perp \ll d$. For sinusoidal optical potentials, we get the explicit condition $4h^2/\lambda\sqrt{2mV_0} \ll 1$. 

In the IL limit, the overlap between the wavefunctions of different layers is negligible and one can assume:
\begin{equation}\label{eq:noverlap}
w_{\alpha m}(z)\,w_{\beta n}(z) \propto \delta_{mn}, \quad \mathrm{for~all}~z\in\mathbb{R},~\mathrm{and}~m, n \in \mathbb{Z}.
\end{equation}
It is straightforward to show using Eqs.~(\ref{eq:J}),~(\ref{eq:V}) and ~(\ref{eq:noverlap}) that the one-body and two-body matrix elements appearing in the second-quantized Hamiltonian assume a much simpler form in this limit:
\begin{eqnarray}
\label{eq:sep1}J_{\alpha\beta}^{mn} &\equiv& \epsilon_{\alpha}\,\delta_{mn}\delta_{\alpha\beta},\\
\label{eq:sep2}\mathcal{V}^{mn;rs}_{\alpha\beta;\gamma\lambda}(\qq) &\equiv& \delta_{mn}\delta_{rs} \mathcal{V}_{\alpha\beta;\gamma\lambda}^{(m-r)}(\qq),
\end{eqnarray}
where $\epsilon_\alpha$ is the zero-point energy of $\alpha$'th subband and is explicitly given by $\hbar \omega_{\mathrm{trap}}(\alpha+1/2)$ in the harmonic trap approximation described above. We note that a more explicit condition for the IL limit in the negligibility of the off-diagonal matrix elements of $J_{\alpha\beta}^{mn}$ and $\mathcal{V}^{mn;rs}_{\alpha\beta;\gamma\lambda}(\qq)$ in the layer indices. Intuitively, Eq.~(\ref{eq:sep1}) and~(\ref{eq:sep2}) imply the absence of inter-layer tunneling and inter-layer exchange interactions, respectively. It is easy to see that the layer index remains a good quantum number in the IL limit in the presence of interactions and this salient feature simplifies the study of multi-layer systems to a great degree.\\

In the next two subsections, we explore the effective microscopic models for single-layer systems ($N_l=1$) and multi-layer cases ($N_l>1$) in some more detail and briefly discuss the features of the normal liquid phase in each case within the Hartree-Fock approximation. 

\subsection{Single-layer systems}\label{sec:HFSL}
The absence of inter-layer attractive interactions in a single-layer system makes it an ideal starting point for the study of the more complicated case of a multi-layer configuration. The physics of single-layer systems is essentially governed by intra-layer repulsive interactions. From an experimental point of view, this limit is achieved either by selectively removing particles from an optical lattice in order to get a single pancake, or by utilizing a strong optical trap instead of an optical lattice. The Hamiltonian takes the following form in this limit:
\begin{eqnarray}\label{eq:HSL}
\mathcal{H}_{\mathrm{SL}} &=& \sum_{\alpha,\kk} \left(|\kk|^2 + E_{\alpha}\right)\,c^{\dagger}_{\alpha,\kk} c^{\phantom{\dagger}}_{\alpha,\kk} + \frac{1}{2A}\,\sum_{\kk_1,\kk_2,\qq}\sum_{\alpha\beta;\gamma\lambda}\nonumber\\
&& 
\mathcal{V}_{\alpha\beta;\gamma\lambda}(\qq)\,c^{\dagger}_{\alpha,\kk_1+\qq} \, c^{\dagger}_{\gamma,\kk_2-\qq} c^{\phantom{\dagger}}_{\lambda,\kk_2} \, c^{\phantom{\dagger}}_{\beta,\kk_1},\nonumber\\
\end{eqnarray} 
where $\mathcal{V}_{\alpha\beta;\gamma\lambda}(\qq) \equiv \mathcal{V}^{(0)}_{\alpha\beta;\gamma\lambda}(\qq)$ are the intra-layer interaction matrix elements. A generating function and explicit expressions for $\mathcal{V}_{\alpha\beta;\gamma\lambda}(\qq)$ can be found in our earlier paper~\cite{Babadi2011}. For trap potentials which are symmetric about their center, the effective inter-subband interaction matrix elements conserve the net subband parity of the scattering particles, i.e. $\mathcal{V}_{\alpha\beta;\gamma\lambda}(\qq) \neq 0$ if $\alpha + \beta + \gamma + \lambda \equiv 0~(\mathrm{mod}~2)$~\cite{Babadi2011}.\\

In an earlier paper, we have studied the self-energy corrections of the single-layer system in the normal liquid phase in Ref.~\cite{Babadi2011} in the self-consistent Hartree-Fock approximation. We do not repeat the analysis here and just mention that besides the usual Hatree-Fock band renormalization, one also finds that the non-interacting subband indices do not remain good quantum numbers in the presence of interactions. A well-defined subband index can still be found after applying an orthogonal transformation that diagonalizes the Hartree-Fock decoupled Hamiltonian. More explicitly, one can define a set of Hartree-Fock fermion operators as: 
\begin{eqnarray}\label{eq:HFop}
\tilde{c}^{\phantom{\dagger}}_{\kk,\alpha}&=&\sum_{\mu} U_{\mu\alpha}(\kk)\,c^{\phantom{\dagger}}_{\kk,\mu},
\end{eqnarray}
such that:
\begin{equation}\label{eq:HSLHF}
\mathcal{H}_{\mathrm{SL}}^{\mathrm{HF}} = \sum_{\alpha,\kk}\tilde{\epsilon}_{\alpha}(\kk)\,\tilde{c}^{\dagger}_{\kk,\alpha}\,\tilde{c}^{\phantom{\dagger}}_{\kk,\alpha},
\end{equation}
where $\tilde{\epsilon}_{\alpha}(\kk)$ are the renormalized energy dispersions of the Hartree-Fock subbands and $\mathcal{H}_{\mathrm{SL}}^{\mathrm{HF}}$ is the Hartree-Fock decoupled Hamiltonian of a single-layer system. The orthogonal transformations appearing in Eq.~(\ref{eq:HFop}), $U_{\mu\alpha}(\kk)$, as well as the renormalized dispersions, $\tilde{\epsilon}_{\alpha}(\kk)$, are found by solving the Hatree-Fock equations (see Ref.~\cite{Babadi2011} for details). The temperature Green's function for Hatree-Fock quasiparticles can be read directly from Eq.~(\ref{eq:HSLHF}):
\begin{eqnarray}\label{eq:tempgreen}
\tilde{\GG}_{\mu\nu}(\kk,i\omega_n) &=& -\int_{0}^{\beta\hbar} \mathrm{d}\tau\,e^{i\omega_n\tau}\,\mathrm{Tr}\big[\hat{\rho}_{\mathrm{SL}}^{\mathrm{HF}}\tilde{c}^{\phantom{\dagger}}_{\kk,\mu}(\tau)\tilde{c}^{\dagger}_{\kk,\nu}(0)\big]\nonumber\\
&=& \frac{\delta_{\mu\nu}}{i\omega_n - \tilde{\xi}_{\mu}(\kk)},
\end{eqnarray}
where $\hat{\rho}_{\mathrm{SL}}^{\mathrm{HF}} = e^{-\beta(\mathcal{H}_{\mathrm{SL}}^{\mathrm{HF}}-\mu\mathcal{N})}/Z_{\mathrm{SL}}$ is the grand canonical equilibrium density matrix and $\tilde{\xi}_{\mu}(\kk) = \tilde{\epsilon}_{\mu}(\kk) - \mu$. The Green's function in the original non-interacting basis can also be found using the inverse of the transformation given in Eqs.~(\ref{eq:HFop}):
\begin{eqnarray}\label{eq:tempgreenorig}
\GG_{\mu\nu}(\kk,i\omega_n) &=& \frac{U_{\mu\lambda}(\kk)\,U_{\nu\lambda}(\kk)}{i\omega_n - \tilde{\xi}_{\lambda}(\kk)}.
\end{eqnarray}
The above expression for the Green's function is found to be useful in evaluating frequency summations later.

\subsection{Multi-layer systems}\label{sec:HFML}
The physics of multi-layer systems is governed by both intra-layer and inter-layer interactions. As we will see later, the interplay of these forces will modify the density-wave instability of the system to a great degree. The Hamiltonian takes the following form in this limit:
\begin{eqnarray}\label{eq:interfinal}
\mathcal{H}_{\mathrm{ML}} &=& \sum_{\kk}\sum_{\alpha,m} \left(|\kk|^2 + E_{\alpha}\right)\,c^{\dagger}_{m\alpha,\kk} c^{\phantom{\dagger}}_{m\alpha,\kk}+\frac{1}{2A}\sum_{\kk_1,\kk_2,\qq} \nonumber\\
&&\sum_{\alpha\beta;\gamma\lambda}\sum_{mr}
\mathcal{V}_{\alpha\beta;\gamma\lambda}^{(m-r)}(\qq)\,c^{\dagger}_{m\alpha,\kk_1+\qq} \, c^{\dagger}_{r\gamma,\kk_2-\qq}\nonumber\\
&&\times\, c^{\phantom{\dagger}}_{r\lambda,\kk_2} \, c^{\phantom{\dagger}}_{m\beta,\kk_1}.
\end{eqnarray}
In contrast to the interaction of particles within each layer, the inter-subband interaction of particles across the layers violate the parity conservation due to the absence of reflection symmetry, i.e. $\mathcal{V}_{\alpha\beta;\gamma\lambda}^{(m-r)}(\qq)$ can still be non-zero if $\alpha + \beta + \gamma + \lambda \equiv 1~(\mathrm{mod}~2)$, provided that $m \neq r$. Explicit expressions for $\mathcal{V}_{\alpha\beta;\gamma\lambda}^{(m,r)}(\qq)$ are provided in Appendix~\ref{sec:app1}.\\

As mentioned in Sec.~\ref{sec:model}, one simplifying aspect of the IL limit is the conservation of the layer indices in the scattering processes (see Eq.~\ref{eq:noverlap}). As a consequence, the normal liquid solution of multi-layer systems in the Hartree-Fock approximation closely resembles that of single-layer systems. The Green's function is found by solving the following Dyson's equation:
\begin{equation}
\parbox{215pt}{\includegraphics{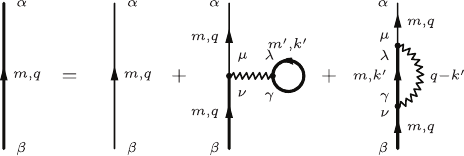}}\quad~,
\end{equation}
where $m$ and $m'$ are layer indices, $q$ is the momentum transfer, the greek letters denote subband indices and thin and thick lines denote bare and dressed Green's functions. The above diagrammatic equation yields:
\begin{eqnarray}\label{eq:mbHFlayer}
\GG_{\alpha\beta;m}(\qq,i\omega_n) & = & \GG^0_{\alpha\beta;m}(\qq,i\omega_n) + \sum_{\mu\nu}G^0_{\alpha\mu;m}(\qq,i\omega_n)\nonumber\\&&\times\,\Sigma^{\star}_{\mu\nu;m}(\qq)\GG_{\nu\beta;m}(\qq,i\omega_n),
\end{eqnarray}
where the non-interacting Green's function, $\GG^0_{\alpha\beta;m}(\qq,i\omega_n)$, is:
\begin{equation}
\GG^0_{\alpha\beta;m}(\qq,i\omega_n) = \frac{\delta_{\alpha\beta}}{i\omega_n - |\kk|^2 - \epsilon_{\alpha} + \mu},
\end{equation}
and the proper self-energy matrix $\Sigma^{\star}_{\mu\nu;m}(\qq)$ is defined as:
\begin{eqnarray}\label{eq:mbsenlayer}
\hspace{-20pt}\Sigma^{\star}_{\mu\nu;m}(\qq) &=& \frac{1}{\beta}\sum_{i\omega'_n;m'}\int \frac{\mathrm{d}^2\kk'}{(2\pi)^2}\sum_{m'}\Big[\mathcal{V}^{(m-m')}_{\mu\nu;\gamma\lambda}(0)\nonumber\\
&&-\mathcal{V}^{(0)}_{\mu\lambda;\gamma\nu}(\qq-\kk')\delta_{mm'}\Big]\GG_{\lambda\gamma;m'}(\kk',i\omega'_n).
\end{eqnarray}
In the homogeneous normal liquid phase, the layers are identical and the Green's functions and self-energy matrices are independent of the layer indices. As a result, the Hatree-Fock equation for multi-layer systems in the IL limit has an identical structure to that of single-layer systems, however, with additional contributions coming from direct inter-layer interactions. Thus, the numerical method described in Ref.~\cite{Babadi2011} can be identically applied to obtain the renormalized subbands of multi-layer systems as well. We refer the reader to Ref.~\cite{Babadi2011} for computational details and suffice it to mention that like single-layer systems, one can again find an orthogonal transformation of the bare fermion operators that diagonalizes the Hartree-Fock-decoupled Hamiltonian. More explicitly, one can define Hartree-Fock quasiparticle operators as:
\begin{equation}\label{eq:HFopsML}
\tilde{c}_{m\alpha,\kk} = \sum_{\mu}U_{\mu\alpha}(\kk)\,c_{m\mu,\kk},
\end{equation} 
such that:
\begin{equation}\label{eq:HMLHF}
\mathcal{H}_{\mathrm{ML}}^{\mathrm{HF}} = \sum_{m,\alpha,\kk}\tilde{\epsilon}_{\alpha}(\kk)\,\tilde{c}^{\dagger}_{m\alpha,\kk}\,\tilde{c}^{\phantom{\dagger}}_{m\alpha,\kk},
\end{equation}
where $\tilde{\epsilon}_{\alpha}(\kk)$ is the renormalized energy dispersion of subband $\alpha$, $U_{\alpha\beta}(\kk)$ is an orthogonal transformation and $\mathcal{H}_{\mathrm{ML}}^{\mathrm{HF}}$ is the Hartree-Fock decoupled Hamiltonian of a multi-layer system. Note that $\tilde{\epsilon}_{\alpha}$ and $U_{\mu\alpha}$ are the same for all layers. The temperature Green's function can be directly read from Eq.~(\ref{eq:HMLHF}) and is formally identical to Eq.~(\ref{eq:tempgreen}) and~(\ref{eq:tempgreenorig}) respectively. The effects of direct inter-layer interactions are implicitly included in the renormalized subband dispersions and orthogonal transformations.

\section{Evaluating the response functions in the TDHF approximation and locating the instabilities of the liquid phase}\label{sec:response}
The instabilities of the liquid phase can be located by calculating various static response functions in the liquid phase. We investigate the instability of the liquid phase toward in-plane and out-of-plane (ripplon) density-wave orders. As a first step, we define the order parameters and their corresponding response functions in more detail in the next subsections.

\subsection{Order parameters}
We define the {\em in-plane projected density operator} of layer $m$ at in-plane coordinate $\xx$ as:
\begin{eqnarray}
\hat{\rho}_m(\xx) & = & \int_{(m-1/2)d}^{(m+1/2)d}\mathrm{d}z\,\psi^{\dagger}(\rr)\psi(\rr)\nonumber\\
& = & \sum_{\alpha,\alpha'}\sum_{m,m'}\sum_{\kk,\kk'} \int_{(m-1/2)d}^{(m+1/2)d}\mathrm{d}z\,w^*_{m\alpha}(z)\,w_{m'\alpha'}(z)\nonumber\\
&&\qquad\qquad\qquad\qquad\quad\times\,e^{-i(\kk-\kk')\cdot\xx}\,c^{\dagger}_{m\alpha,\kk}c^{\phantom{\dagger}}_{m'\alpha',\kk'}\nonumber\\
&\approx&\sum_{\alpha}\sum_{\kk,\kk'}e^{-i(\kk-\kk')\cdot\xx}\,c^{\dagger}_{m\alpha,\kk}c^{\phantom{\dagger}}_{m\alpha,\kk'},
\end{eqnarray}
where we have adopted the IL approximation in the last line. In the normal phase, $\langle\hat{\rho}_m(\xx)\rangle$ is constant and independent of $\xx$. The in-plane density-wave instability is characterized by appearance of (quasi-)periodic spatial modulations in $\langle\hat{\rho}_m(\xx)\rangle$ (see Fig.~\ref{fig:phases}c).\\

We define the {\em $\alpha\beta$-ripplon operator} of layer $m$ at in-plane coordinate $\xx$ as:
\begin{eqnarray}
\hat{R}^{\alpha\beta}_m(\xx) & = & \frac{1}{2}\int_{(m-1/2)d}^{(m+1/2)d}\mathrm{d}z\,\left(\psi^{\dagger}_{\alpha}(\rr)\psi^{\phantom{\dagger}}_{\beta}(\rr) + \mathrm{h.c.}\right)\nonumber\\
&\approx&\frac{1}{2}\sum_{\kk,\kk'}e^{-i(\kk-\kk')\cdot\xx}\,\left(c^{\dagger}_{m\alpha,\kk}c^{\phantom{\dagger}}_{m\beta,\kk'} + c^{\dagger}_{m\beta,\kk}c^{\phantom{\dagger}}_{m\alpha,\kk'}\right).\nonumber\\
\end{eqnarray}
Again, we have adopted the IL approximation in the last line. In the normal phase, $\langle\hat{S}^{\alpha\beta}_m(\xx)\rangle = 0$ for $\alpha\neq\beta$. The $\alpha\beta$-ripplon instability is characterized by growth of (quasi-)periodic spatial modulations in $\langle\hat{S}^{\alpha\beta}_m(\xx)\rangle$ and absence of any instability in the in-plane projected density~(see Fig.~\ref{fig:phases}b). When both in-plane and out-of-plane symmetries are broken, we refer to the case as the {\em zigzag} instability (see Fig.~\ref{fig:phases}d).

\subsection{Evaluation of the response functions}
We evaluate the response functions in the imaginary time formalism and find the real-time response functions by analytic continuation. The imaginary-time in-plane projected density-density response function is defined as:
\begin{equation}
\chi_{\mathrm{dd}}^{(m-m')}(\xx \tau; \xx' \tau') = -\mathrm{Tr} \left\{ \hat{\rho}_\mathrm{G} \mathrm{T}_\tau \left[ \hat{\rho}_m(\xx \tau) \hat{\rho}_{m'}(\xx' \tau') \right] \right\},
\end{equation}
where $\hat{\rho}_\mathrm{G}~=~e^{-\beta(\mathcal{H}-\mu \mathcal{N})}/Z_{\mathrm{G}}$ is the grand canonical weighting operator. The imagiary-time ripplon-ripplon response function is defined as:
\begin{equation}
\chi_{\alpha\beta}^{(m-m')}(\xx \tau; \xx' \tau') = -\mathrm{Tr}\,\{ \hat{\rho}_\mathrm{G} \mathrm{T}_\tau [ \hat{S}^{\alpha\beta}_m(\xx \tau) \hat{S}^{\alpha\beta}_{m'}(\xx' \tau') ] \}.
\end{equation}
The space and time translation invariance of the Hamiltonian and the normal phase imply that the response functions are functions of $\xx-\xx'$ and $\tau-\tau'$ and therefore, it is easier to express them in terms of transferred momentum $\qq$ and bosonic Matsubara frequencies $i\nu_n$. Also, both of the density-desity and ripplon-ripplon response functions can be expressed in terms of polarization insertions as follows:
\begin{equation}\label{eq:chiddpol}
\chi_{\mathrm{dd}}^{(m-m')}(\qq,i\nu_n) = \sum_{\alpha,\beta}\Pi^{(m-m')}_{\alpha\alpha;\beta\beta}(\qq,i\nu_n),
\end{equation}
and:
\begin{multline}\label{eq:chiabpol}
\chi_{\alpha\beta}^{(m-m')}(\qq,i\nu_n) = \frac{1}{4}\Big[\Pi^{(m-m')}_{\alpha\beta;\alpha\beta}(\qq,i\nu_n) + \Pi^{(m-m')}_{\beta\alpha;\alpha\beta}(\qq,i\nu_n)\\
+\,\Pi^{(m-m')}_{\alpha\beta;\beta\alpha}(\qq,i\nu_n) + \Pi^{(m-m')}_{\beta\alpha;\beta\alpha}(\qq,i\nu_n)\Big],
\end{multline}
where the polarization insertion is defined as:
\begin{multline}\label{eq:poldefn}
\Pi^{(m-m')}_{\alpha\beta;\gamma\lambda}(\qq,i\nu_n) = \frac{1}{A}\sum_{\kk,\kk'}\Pi^{(m-m')}_{\alpha\beta;\gamma\lambda}(\qq,i\nu_n; \kk,\kk'),
\end{multline}
and:
\begin{multline}\label{eq:poldefn2}
\Pi^{(m-m')}_{\alpha\beta;\gamma\lambda}(\qq,i\nu_n;\kk,\kk') =
\int_0^{\beta} d\tau \, e^{i\nu_n\tau} \\ \Big\langle c^{\dagger}_{m\alpha,\kk+\qq}(\tau) c^{\phantom{\dagger}}_{m\beta,\kk}(\tau) c^{\dagger}_{m'\gamma,\kk'-\qq}(0) c^{\phantom{\dagger}}_{m'\lambda,\kk'}(0) \Big\rangle_{\mathrm{con.}}.
\end{multline}
Only diagrams with connected external vertices must be considered in Eq.~(\ref{eq:poldefn2}).\\

The TDHF approximation for the polarization insertion amounts to summing ladder and ring diagrams to all orders~\cite{NozieresPines,Baym1961}. Although we are only interested in the static limit in this study, i.e. $i\nu_n \rightarrow i0^+$, we will only take this limit at the end of the derivation for generality. A typical term contributing to $\Pi^{(m-m')}_{\alpha\beta;\gamma\lambda}$ in the TDHF approximation consists of one or more bubble diagrams, possibly with ladder-type vertex corrections, connected to each other by interaction lines:
\begin{eqnarray}\label{eq:TDHFtypical}\nonumber
&&\Pi^{(m-m')}_{\alpha\beta;\gamma\lambda}=\ldots \nonumber\\
&&\qquad+\quad
\parbox{195pt}{\includegraphics{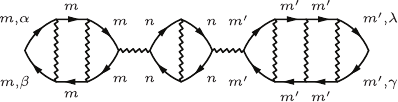}} \nonumber\\
  &&\qquad+\quad\ldots\,
\end{eqnarray}
Since the layer index in conserved on each interaction vertex, it is easy to see that the particle and hole lines appearing in an irreducible polarization diagram (bubble) carry the same layer index. Thus, the vertex corrections are only due to the intra-layer interactions. The homogeneity of the normal phase also implies that the bubble diagrams are independent of the layer indices. Thus, we can carry out the summation in two steps: first, we evaluate the irreducible polarizations by summing the ladder-type vertex corrections to all orders. Next, we calculate the full polarization by connecting the bubbles with interaction lines. 

Let $\Pi^{\star}_{\alpha\beta;\gamma\lambda}$ be the irreducible intera-layer particle-hole propagator with ladder-like interactions summed to all orders. $\Pi^{\star}_{\alpha\beta;\gamma\lambda}$ can be found by solving the following Bethe-Salpeter equation:
\begin{eqnarray}\label{eq:Iladder}
&&\hspace{-15pt}\Pi^{\star}_{\alpha\beta;\gamma\lambda}(\qq,i\nu_n;\kk_1,\kk_2)=
\parbox{70pt}{\includegraphics{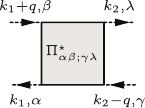}}\nonumber\\&&=\,
\parbox{190pt}{\includegraphics{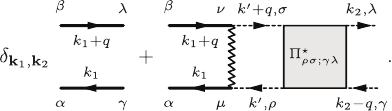}}
\end{eqnarray}
The diagrammatic equation yields the following integral equation:
\begin{multline}\label{eq:integeq1}
\Pi^{\star}_{\alpha\beta;\gamma\lambda}(\qq,i\nu_n;\kk_1) = \Pi^{(0)}_{\alpha\beta;\gamma\lambda}(\kk_1,\qq,i\nu_n)\\-\Pi^{(0)}_{\alpha\beta;\mu\nu}(\qq,i\nu_n;\kk_1)\,\int\frac{\mathrm{d}^2\kk'}{(2\pi)^2}\,\mathcal{V}_{\sigma\nu;\mu\rho}(\kk'-\kk_1)\\
\times\,\Pi^{\star}_{\rho\sigma;\gamma\lambda}(\qq,i\nu_n;\kk'),
\end{multline}
where we have summed both sides over $\kk_2$. Summation over repeated indices is assumed throughout. $\Pi^{(0)}_{\alpha\beta;\mu\nu}(\qq,i\nu_n;\kk)$ is the bare particle-hole propagator:
\begin{multline}
\Pi^{(0)}_{\alpha\beta;\gamma\lambda}(\qq,i\nu_n;\kk) = \\
\frac{1}{\beta}\sum_{i\omega_n}\mathcal{G}_{\lambda\beta}(\kk+\qq,i\omega_n+i\nu_n)\mathcal{G}_{\alpha\gamma}(\kk,i\omega_n)= \\
U_{\alpha\alpha'}(\kk)U_{\beta\beta'}(\kk+\qq)U_{\gamma\gamma'}(\kk)U_{\lambda\lambda'}(\kk+\qq)\\
\times\,\delta_{\beta'\lambda'}\delta_{\alpha'\gamma'}\,\frac{n^{F}\big(\tilde{\xi}_{\kk,\alpha'}\big)-n^{F}\big(\tilde{\xi}_{\kk+\qq,\beta'}\big)}{i\nu_n-\big(\tilde{\xi}_{\kk+\qq,\beta'}-\tilde{\xi}_{\kk,\alpha'}\big)}
\end{multline}
The irreducible polarization diagram, $\Pi^{\star}_{\alpha\beta;\gamma\lambda}(\qq,i\nu_n)$, is found by summing $\Pi^{\star}_{\alpha\beta;\gamma\lambda}(\qq,i\nu_n;\kk_1,\kk_2)$ over $\kk_1$ and $\kk_2$. The summation over $\kk_2$ is trivial and is already done in Eq.~(\ref{eq:integeq1}). The summation over $\kk_1$, however, may only be done once the solution of the integral equation is known. We solve the integral equation numerically. The details of the numerical procedure is provided in Appendix.~\ref{app:num}.\\

Once $\Pi^{\star}_{\alpha\beta;\gamma\lambda}(\qq,i\nu_n)$ is evaluated, the full polarization can be easily obtained by summing the ring diagrams to all orders. We note that the interaction lines connecting the irreducible polarizations may have vertices belonging to different layers (see Eq.~\ref{eq:TDHFtypical}). The following Dyson's equation yields the summation ring diagrams to all orders:
\begin{eqnarray}
&&\hspace{-20pt}\Pi^{(m-m')}_{\alpha\beta;\gamma\lambda}(\qq,i\nu_n) \equiv
\parbox{60pt}{\includegraphics{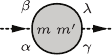}} \nonumber\\&&\qquad= \delta_{mm'}~\parbox{60pt}{\includegraphics{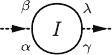}} \nonumber\\
&&\quad\qquad+\sum_{\mu\nu;\rho\sigma}\sum_n~\parbox{110pt}{\includegraphics{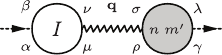}},
\end{eqnarray}
where the blank and filled circles denote irreducible and full polarizations, respectively. The above diagrammatic equation yields the following linear system of equations:
\begin{eqnarray}\label{eq:dysonlayer}
\hspace{-15pt}\Pi^{(m-m')}_{\alpha\beta;\gamma\lambda} &=& \delta_{mm'}\Pi^{\star}_{\alpha\beta;\gamma\lambda} \nonumber\\&&+ \sum_{n=-N_l+1}^{N_l-1}\,\sum_{\mu\nu;\rho\sigma}\Pi^{\star}_{\alpha\beta;\mu\nu}\,\mathcal{V}^{(m-n)}_{\mu\nu;\rho\sigma}\,\Pi^{(n-m')}_{\rho\sigma;\gamma\lambda},
\end{eqnarray}
where we have dropped the common argument $(\qq,i\nu_n)$ for brevity. Since we assumed periodic boundary conditions along the $z$-axis, $\Pi^{(m-m')}_{\alpha\beta;\gamma\lambda}$ is a periodic function of $m-m'$ and the above equation can be most conveniently solved by going from the layer index to transverse momentum representation. We define:
\begin{equation}\label{eq:fourier}
\tilde{\Pi}^{(k_z)}_{\alpha\beta;\gamma\lambda}(\qq,i\nu_n) = \sum_{m=0}^{N_l-1} e^{-i k_z m d}\,\Pi^{(m)}_{\alpha\beta;\gamma\lambda}(\qq,i\nu_n),
\end{equation}
where $k_z = 2\pi n / L$ for $n=0, 1, \ldots, N_l-1$. Plugging Eq.~(\ref{eq:fourier}) into Eq.~(\ref{eq:dysonlayer}), we get:
\begin{equation}\label{eq:dysonlayer3}
\tilde{\Pi}^{(k_z)}_{\alpha\beta;\gamma\lambda} = \Pi^{\star}_{\alpha\beta;\gamma\lambda} + \sum_{\mu\nu;\rho\sigma} \Pi^{\star}_{\alpha\beta;\mu\nu}\,\tilde{\mathcal{V}}^{(k_z)}_{\mu\nu;\rho\sigma}\,\tilde{\Pi}^{(k_z)}_{\rho\sigma;\gamma\lambda},
\end{equation}
where:
\begin{equation}\label{eq:Vk}
\tilde{\mathcal{V}}^{(k_z)}_{\mu\nu;\rho\sigma}(\qq) = \sum_{n=-N_l+1}^{N_l-1} e^{-i k_z n d}\,\mathcal{V}^{(n)}_{\mu\nu;\rho\sigma}(\qq).
\end{equation}
The transverse modes with different $k_z$ are decoupled in Eq.~(\ref{eq:dysonlayer3}) and the problem reduces to solving a linear system in the subband indices for each $k_z$. The response functions can also be expressed conveniently in the transverse momentum basis using Eqs.~(\ref{eq:chiddpol}) and~(\ref{eq:chiabpol}):
\begin{align}\label{eq:chiddpolk}
\chi_{\mathrm{dd}}^{(k_z)}(\qq,i\nu_n) =&\sum_{\alpha,\beta}\Pi^{(k_z)}_{\alpha\alpha;\beta\beta}(\qq,i\nu_n),\\
\label{eq:chiabpolk}
\chi_{\alpha\beta}^{(k_z)}(\qq,i\nu_n) =&\frac{1}{4}\Big[\Pi^{(k_z)}_{\alpha\beta;\alpha\beta}(\qq,i\nu_n) + \Pi^{(k_z)}_{\beta\alpha;\alpha\beta}(\qq,i\nu_n)\nonumber\\
&+\,\Pi^{(k_z)}_{\alpha\beta;\beta\alpha}(\qq,i\nu_n) + \Pi^{(k_z)}_{\beta\alpha;\beta\alpha}(\qq,i\nu_n)\Big].
\end{align}

Before embarking on evaluating the response function using the described formalism, we find it worthwhile to briefly study the direct consequences of the coupling between in-plane and out-of-plane modes. Understanding the coupling between various density ordering modes guides us in predicting which modes go unstable simultaneously and which modes may remain stable once the liquid phase becomes unstable.

It is straightforward to establish that all in-plane density fluctuations (corresponding to polarization diagrams such as $\Pi_{00;00}$, $\Pi_{00;11}$, $\Pi_{11;11}$, etc.) are coupled to each other due to the existence of interaction matrix elements $\mathcal{V}_{00;11}$ and such. Therefore, the in-plane density wave modes go unstable together and contribute to the formation of an inhomogeneous case. In particular, coexistence of liquid phase in one subband and a density ordered phase in another subband is impossible.

As mentioned in Sec.~\ref{sec:HFSL}, the inter-subband interactions conserve the net of parity of the interacting quasiparticles in the single-layer case. As a consequence, there is no coupling between in-plane density fluctuations and odd ripplons (corresponding to polarization diagrams such as $\Pi_{01;01}$, $\Pi_{01;10}$, etc) due to the absence of interaction matrix elements $\mathcal{V}_{00;01}$ and such. For instance, starting from the $\mathrm{N}_1$ phase, it is possible to reach a density ordered phase with no accompanying out-of-plane order.

In multi-layer systems ($N_l>1$), the situation can be different. As mentioned in Sec.~\ref{sec:HFML}, the inter-subband interactions between quasiparticles of different layers violate the parity conservation. Using the results of Appendix~\ref{sec:app1}, one easily finds that the parity violating interaction matrix elements are odd under the inversion of layer indices, i.e. $\mathcal{V}^{(m-m')}_{\alpha\beta;\gamma\lambda}(\qq) = (-1)^{P}\,\mathcal{V}^{(m'-m)}_{\alpha\beta;\gamma\lambda}(\qq)$, where $P = (\alpha+\beta+\gamma+\lambda)~\mathrm{mod}~2$. Using this property, the inter-layer interactions in the transverse momentum basis, Eq.~(\ref{eq:Vk}), can be expressed in a more useful form:
\begin{equation}\label{eq:Vk2}
\tilde{\mathcal{V}}^{(k_z)}_{\alpha\beta;\gamma\lambda}(\qq) = \left\{
\begin{tabular}{l}
$\displaystyle\mathcal{V}_{\alpha\beta;\gamma\lambda}(\qq) \, + \, 2\sum_{n=1}^{N_l-1} \cos(k_z n d)\,\mathcal{V}^{(n)}_{\alpha\beta;\gamma\lambda}(\qq)$\\
\qquad\qquad\qquad\qquad\qquad\qquad\quad~$\mathrm{if~}P=0$,\\
$\displaystyle-2 i \sum_{n=1}^{N_l-1} \sin(k_z n d)\,\mathcal{V}^{(n)}_{\alpha\beta;\gamma\lambda}(\qq)$\\
\qquad\qquad\qquad\qquad\qquad\qquad\quad~$\mathrm{if~}P=1$,
\end{tabular}\right.
\end{equation}
Clearly, the parity violating matrix elements ($P=1$) are non-vanishing only if $k_z \neq 0$. Therefore, in multi-layer systems, density waves and odd ripplons are coupled at finite transverse momenta. Thus, if the first mode that goes unstable has a finite transverse momentum, the resulting ordered phase breaks both in-plane translation symmetry and $\mathbb{Z}_2$ reflection symmetry.\\

We note that once the leading instability is found, the study of subsequent instabilities must be done with a word of caution. The leading instability modifies the initial state, either by producing short range correlations or breaking a symmetry. Even if the new state can be described well at the mean-field level, the Green's function and the response functions must be recalculated in the new state. This requirement in turn modifies the nature and/or order of the subsequent instabilities. We only study the leading instability of the liquid phase in this paper and leave the study of subsequent transitions within the density-ordered phase for future works.

\section{The Mean-field instability diagram of the liquid phase}\label{sec:res}
Due to the complexity of the formalism described in the previous section, obtaining analytical expressions for the response functions in the TDHF approximation is a formidable task without resorting to further approximations. The most involved part of the calculation is solving the Bethe-Salpeter integral equation that represents the effects of intra-layer exchange interactions. Here, we present the results obtained by exact numerical calculations based on the procedure outlined in the previous section. The numerical procedure is described in Appendix~\ref{app:num} in detail. Later, we will employ further simplifying approximations in order to obtain tractable analytical expressions that guide us in interpreting the results.\\

We are interested in finding the boundary of the stability of the liquid phase and the characteristics of the mode that drives the instability, as a function of tunable parameters of the system. For a fixed number of layers $N_l$, inter-layer separation $d$ and temperature $T$, there remains two tunable dimensionless parameters: the dipolar interaction strength, $r_d$ (see Eq.~\ref{eq:rd}), and the ratio of the transverse confinement width and the mean inter-particle distance, $\sqrt{n}a_\perp$. The limits $\sqrt{n}a_{\perp} \ll 1$, $\sqrt{n}a_{\perp} \sim 1$ and $\sqrt{n}a_{\perp} \gg 1$ correspond to the two-, quasi-two- and three-dimensional regimes respectively. The IL limit is achieved for $a_{\perp}/d \ll 1$. In the study of multi-layer systems, we restrict our parameters to $\sqrt{n}d \gtrsim 1$, i.e. the high density limit (with respective to the layer spacing), in which both IL limit and quasi-two-dimensionality can be approximately achieved. 

We locate the instability boundaries of the liquid phase using a divide and conquer method. For each $a_\perp$, we first locate $r_{d,\mathrm{L}}$ and $r_{d,\mathrm{H}}$ such that all response functions are regular and smooth for $r_{d,\mathrm{L}}$ and at least one mode is unstable at $r_{d,\mathrm{H}}$. The instability appears as a zero crossing in the inverse of some response function. Once a lower and upper limit is found for the critical $r_d$, the exact location of the phase boundary is determined by successive bisection of this interval.

In order to simplify our analysis, we confine our attention to the low temperatures, where thermal fluctuations are negligible compared to the quantum fluctuations. We set $T = 0.02\,T_F^{(0)}$, where $T_F^{(0)} = 2 \pi n\hbar^2 / m k_B$ is the Fermi temperature of a two-dimensional free Fermi gas at the same density. We will later show that the chosen small temperature is high enough to suppress the inter-layer superfluid transition~\cite{Potter2010,Pikovski2010,Baranov2011} in all of the studied multi-layer configurations. 

\subsection{Instabilities of single-layer systems}
We have studied the properties of the liquid phase of single-layer systems in an earlier paper~\cite{Babadi2011}. In brief, when $\sqrt{n}a_{\perp} \ll 1$, the energy gap between the subbands is much larger than the Fermi energy and the system is effectively two-dimensional, i.e. only the lowest subband ($\alpha=0$) is filled. Upon relaxing the trap, i.e. $\sqrt{n}a_{\perp} \sim 1$, the subband gap is reduced and higher subbands will be filled. We denote a normal liquid phase having up to $j$'th subband filled by $\mathrm{N}_j$. The Fermi surface of a system in the $\mathrm{N_j}$ phase consists of $j+1$ circles, characterized by their radii $k_{F,0},~k_{F,1},~\ldots~,~k_{F,j}$. In analogy to quasi-two-dimensional electron gas, we expect to get $j+1$ peaks in static density-density response function vs. momentum $q$ at $q \approx 2k_{F,0},~q \approx 2k_{F,1},~\ldots~,~q \approx 2k_{F,j}$, corresponding to softened particle-hole excitations arising from opposite poles of the Fermi surface of each subband. We also expect to get a single peak at $q \approx k_{F,\alpha}+k_{F,\beta}$ in the $\alpha\beta$-ripplon response function, again, analogous to SDW softening in electron gas~\cite{Overhauser1962}.\\

\begin{figure}[t!]
\includegraphics[width=7cm]{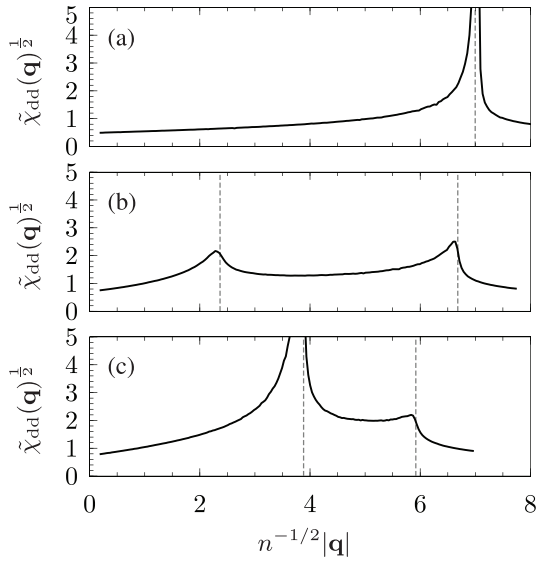}
\caption{Static density-density response function of a single-layer system in the normal phase. $\tilde{\chi}_{\mathrm{dd}} \equiv 2\pi\hbar^2\chi_{\mathrm{dd}}/m$ and the dashed lines denote $q=2k_{F,j}$,~ $j=0,1$. (a) $\mathrm{N}_0$ phase [$r_d$ = 1.0, $\sqrt{n}a_{\perp}$ = 0.15] (b) $\mathrm{N}_1$ phase [$r_d$ = 1.35, $\sqrt{n}a_{\perp}$ = 0.25] (c) $\mathrm{N}_1$ phase [$r_d$ = 1.35, $\sqrt{n}a_{\perp}$ = 0.35]. }
\label{fig:susc-nn-SL}
\end{figure}

Fig.~\ref{fig:susc-nn-SL} shows the static density-density response function in $\mathrm{N}_0$ (top plot) and $\mathrm{N}_1$ (middle and bottom plots) phases. In the $\mathrm{N}_0$ phase, we only get a single peak corresponding to the softened density-wave mode at $q \approx 2 k_{F,0}$. The middle and bottom plots ($\mathrm{N}_1$) correspond to low and high population of the first excited subband. It is noticed that in the middle plot, the $q \approx 2k_{F,0}$ mode is more enhanced compared to $q \approx 2k_{F,1}$ mode. The scenario is reversed, however, as the population of the first subband is increased beyond a certain threshold. Thus, we generally expect $q \approx 2k_{F,0}$ to be the first mode to go unstable in the $\mathrm{N}_0$ phase, while we expect a switching from $q \approx 2k_{F,0}$ to $q \approx 2k_{F,1}$ in the $\mathrm{N}_1$ phase.

\begin{figure}[t!]
\includegraphics[width=7cm]{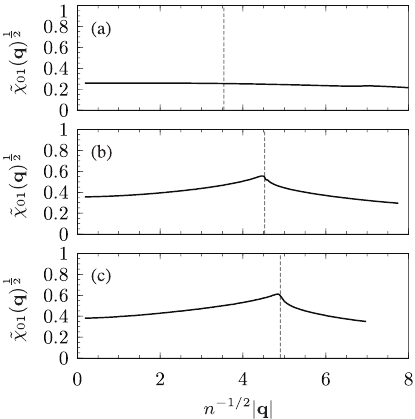}
\caption{Static 01-ripplon response function of a single-layer system in the normal phase. $\tilde{\chi}_{\mathrm{01}} \equiv 2\pi\hbar^2\chi_{\mathrm{01}}/m$ and the dashed lines denote $q=k_{F,0} + k_{F,1}$. (a) $\mathrm{N}_0$ phase [$r_d$ = 1.0, $\sqrt{n}a_{\perp}$ = 0.15] (b) $\mathrm{N}_1$ phase [$r_d$ = 1.35, $\sqrt{n}a_{\perp}$ = 0.25] (c) $\mathrm{N}_1$ phase [$r_d$ = 1.35, $\sqrt{n}a_{\perp}$ = 0.35].}
\label{fig:susc-01-SL}
\end{figure}

\begin{figure}[h!]
\includegraphics[width=7cm]{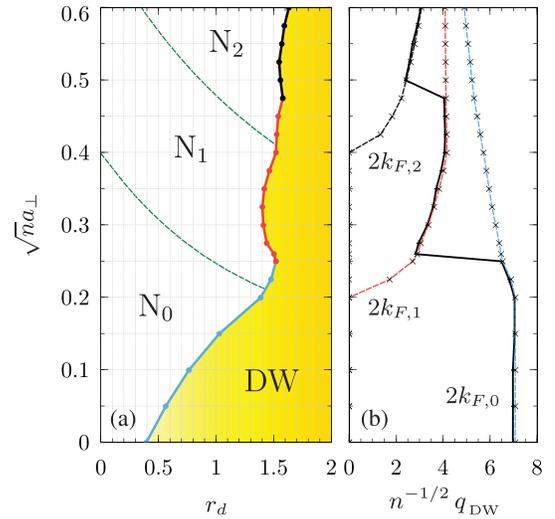}
\caption{(Color online) (a) The phase diagram of quasi-two-dimensional dipolar fermions in a single-layer configuration. The green dashed lines show the boundary between different multi-subband normal phases ($\mathrm{N}_0$, $\mathrm{N}_1$, $\ldots$), the yellow shaded region is a density ordered phase and the thick lines on the $\mathrm{N}$-$\mathrm{DW}$ boundary indicate the unstable wave-vector, $q=2k_{F,0}$ (lower segment, sky blue), $q=2k_{F,1}$ (middle segment, red) and $q=2k_{F,2}$ (upper segment, black). (b) The variation of unstable wave-vector along the $\mathrm{N}$$-$$\mathrm{DW}$ boundary (black line). The blue, red and black dashed lines show twice theFermi momentum of the zeroth, first and second subband on the boundary.}
\label{fig:phase-diag-SL}
\end{figure}

Fig.~\ref{fig:susc-01-SL} shows the static 01-ripplon response function for the same configurations as in Fig.~\ref{fig:susc-nn-SL}. A slight enhancement of the 01-ripplon mode at $q=k_{F,0}+k_{F,1}$ is noticed in $\mathrm{N}_1$ phase, however, the peaks are less pronounced than the peaks of the density-density response function. This result can be understood in light of the stronger intra-subband vs. inter-subband repulsion, the latter being weaker due to contributions from attractive head-to-tail dipole-dipole interactions. Therefore, we generally expect the density-wave instability to preceed the ripplon instability.

Fig.~\ref{fig:phase-diag-SL} shows the instability phase diagram of a single-layer system as a function of $r_d$ and $\sqrt{n}a_\perp$. As we speculated before, we find that the density-wave instability preceeds the ripplon instability in the studied range of parameters. Therefore, the ripplon instability may only appear in the density ordered phase and form a zigzag phase (see Fig.~\ref{fig:phases}d). The plot next to the phase diagram in Fig.~\ref{fig:phase-diag-SL} shows the wave-vector of the unstable mode on the $\mathrm{N}$-$\mathrm{DW}$ boundary. The switching of unstable mode in the $\mathrm{N}_1$ can also be clearly seen: the density ordering wave-vector of a system in the $\mathrm{N}_1$ liquid phase is $q = 2k_{F,0}$ for $\sqrt{n}a_{\perp} < 0.25$, however, it discontinuously jumps to $q = 2k_{F,1}$ for $\sqrt{n}a_{\perp} > 0.25$. The same behavior is observed in the $\mathrm{N}_2$ phase as well. We will investigate this behavior in Sec.~\ref{sec:RPA}. 

\subsection{Instabilities of multi-layer systems}
As mentioned in Sec.~\ref{sec:HFML}, the normal phase of multi-layer systems is very similar to single-layer systems in the IL limit, the only difference being existence of a mean-field shift of the subband energies due to direct inter-layer interactions. The inter-layer interactions, however, can drammatically affect the density wave fluctuations. In particular, one expects a more pronounced enhancement of both density wave and ripplon fluctuations. Analgous to the single-layer case, starting from the $\mathrm{N}_j$ phase, we again expect to see $j+1$ peaks in the static density-density response functions at $q=2k_{F,0}$, $q=2k_{F,1}$, $\ldots$, $q=2k_{F,j}$ and a peak at $q=k_{F,\alpha}+k_{F,\beta}$ in the $\alpha\beta$-ripplon response function. The coupling between density-wave and ripplon modes at finite transverse momenta results in the mixing of these peaks such that traces of density wave peaks can be noticed in the ripplon response function and vice versa. In the following discussions, we keep the number of layers constant, $N_l=50$, which is in the order of the typical number achievable in the experiments.\\

\begin{figure}[t!]
\includegraphics[width=7cm]{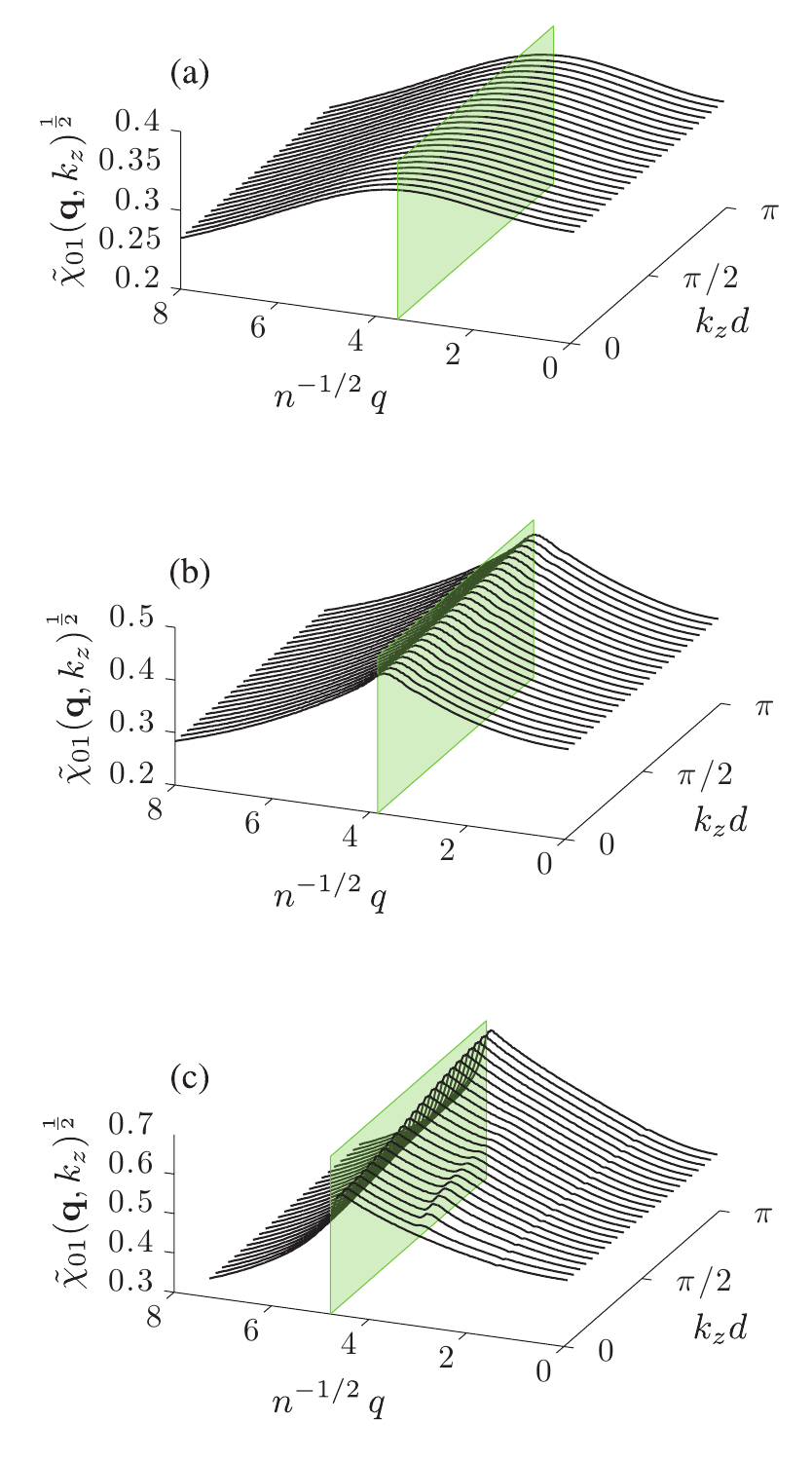}
\caption{(Color online) The static density-density response function of a multi-layer system ($\sqrt{n}d = 1.25$, $N_l=50$) in the normal phase. $\tilde{\chi}_{\mathrm{dd}} \equiv 2\pi\hbar^2\chi_{\mathrm{dd}}/m$ and the blue and red planes denote $q=2k_{F,0}$ and $q=2k_{F,0}$ respectively. (a) $\mathrm{N}_0$ phase [$r_d$ = 1.265, $\sqrt{n}a_{\perp}$ = 0.20] (b) $\mathrm{N}_1$ phase [$r_d$ = 1.255, $\sqrt{n}a_{\perp}$ = 0.22] (c) $\mathrm{N}_1$ phase [$r_d$ = 0.845, $\sqrt{n}a_{\perp}$ = 0.36]. In all plots, it is noticed that $k_z=0$ modes experience the most softening.}
\label{fig:susc-nn-ML}
\end{figure}

\begin{figure}[t!]
\includegraphics[width=7cm]{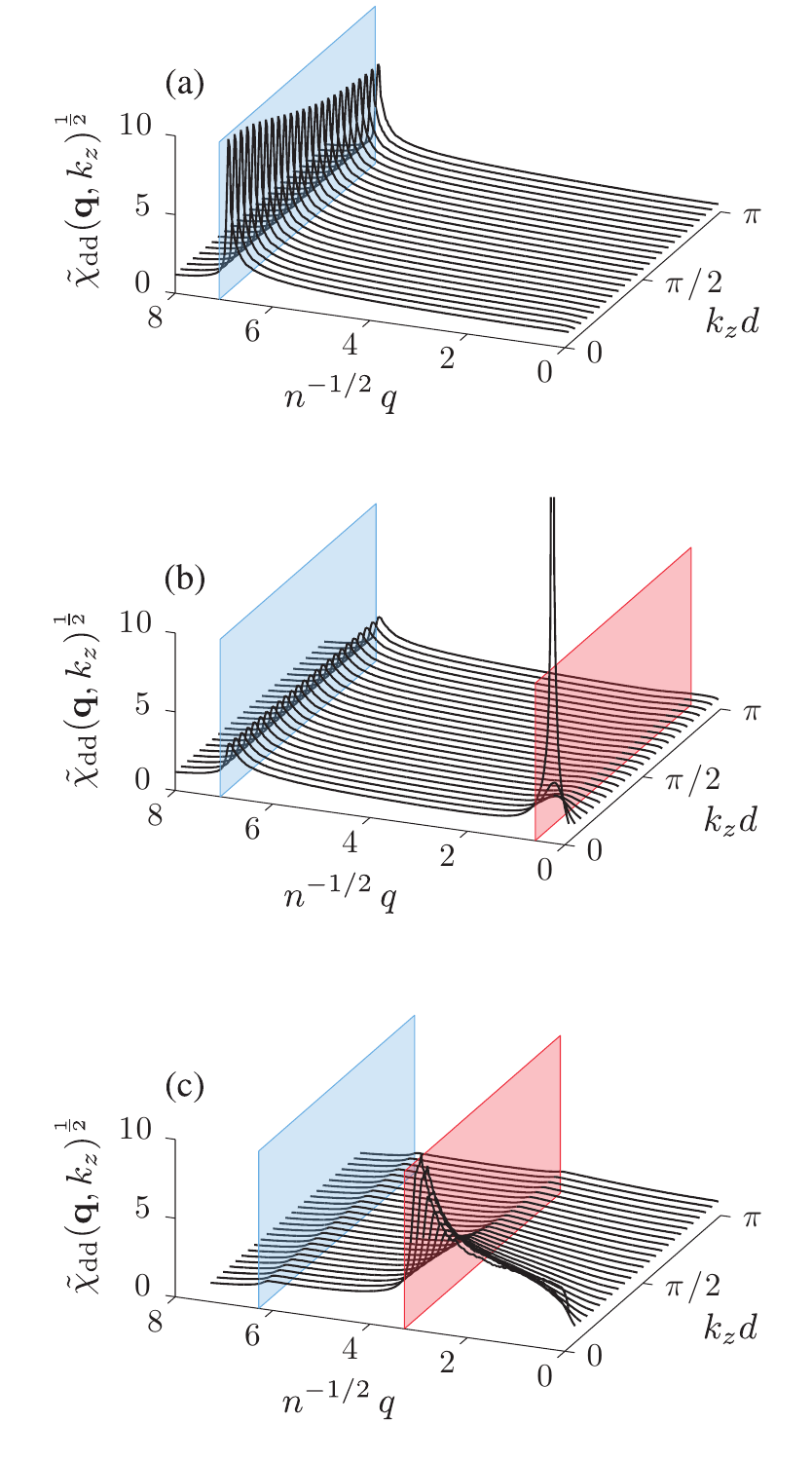}
\caption{(Color online) The static 01-ripplon response function of a multi-layer system ($\sqrt{n}d = 1.25$, $N_l=50$) in the normal phase. $\tilde{\chi}_{\mathrm{01}} \equiv 2\pi\hbar^2\chi_{\mathrm{01}}/m$ and the green planes denote $q=k_{F,0}+k_{F,1}$. (a) $\mathrm{N}_0$ phase [$r_d$ = 1.265, $\sqrt{n}a_{\perp}$ = 0.20] (b) $\mathrm{N}_1$ phase [$r_d$ = 1.255, $\sqrt{n}a_{\perp}$ = 0.22] (c) $\mathrm{N}_1$ phase [$r_d$ = 0.845, $\sqrt{n}a_{\perp}$ = 0.36]. It is noticed that $k_z d=\pi$ modes experience the maximum enhancements due to interactions. The peaks at $n^{-1/2}q \simeq 3$ for finite $k_z$ in plot (c) correspond to the softened density-waves which are coupled to the 01-ripplons through parity-violating inter-layer interactions.}
\label{fig:susc-01-ML}
\end{figure}

\begin{figure}[h!]
\includegraphics[width=7cm]{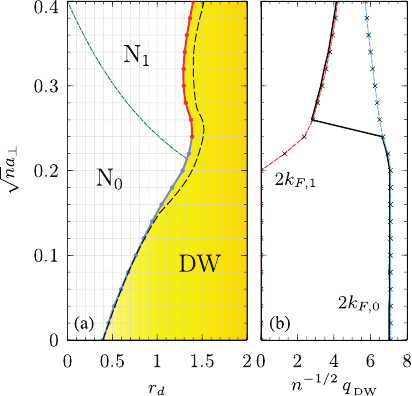}
\caption{(Color online) The phase diagram of quasi-two-dimensional dipolar fermions in a multi-layer configuration ($\sqrt{n}d = 2$, $N_l = 50$). The black dashed line is the $\mathrm{N}$-$\mathrm{DW}$ boundary in the single-layer configuration (refer to to caption of Fig.~\ref{fig:phase-diag-SL} for the description of the lines and symbols).}
\label{fig:phase-diag-ML-2.0}
\end{figure}

\begin{figure}[h!]
\includegraphics[width=7cm]{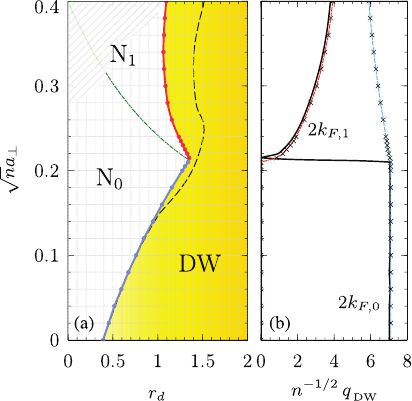}
\caption{(Color online) The phase diagram of quasi-two-dimensional dipolar fermions in a multi-layer configuration ($\sqrt{n}d = 1.5$, $N_l = 50$). The black dashed line is the $\mathrm{N}$-$\mathrm{DW}$ boundary in the single-layer configuration (refer to to caption of Fig.~\ref{fig:phase-diag-SL} for the description of the lines and symbols). The hatched region is where the IL limit is not applicable (Ref. to Sec.~\ref{sec:model}).}
\label{fig:phase-diag-ML-1.5}
\end{figure}

\begin{figure}[h!]
\includegraphics[width=7cm]{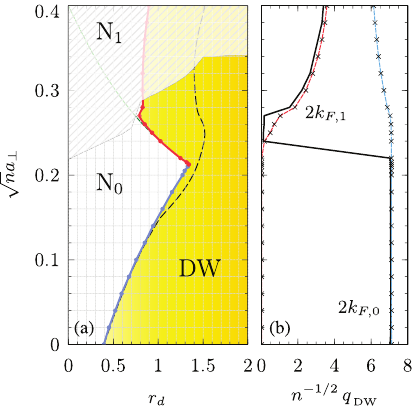}
\caption{(Color online) The phase diagram of quasi-two-dimensional dipolar fermions in a multi-layer configuration ($\sqrt{n}d = 1.25$, $N_l = 50$). The black dashed line is the $\mathrm{N}$-$\mathrm{DW}$ boundary in the single-layer configuration (refer to to caption of Fig.~\ref{fig:phase-diag-SL} for the description of the lines and symbols). The hatched region is where the IL limit is not applicable (Ref. to Sec.~\ref{sec:model}).}
\label{fig:phase-diag-ML-1.25}
\end{figure}

Figs.~\ref{fig:susc-nn-ML} and~\ref{fig:susc-01-ML} show the static density-density and 01-ripplon response functions evaluated for three different points in the normal phase: the (a) plots correspond to a point in the $\mathrm{N}_0$ phase, the (b) plots are in $\mathrm{N}_1$ phase with a small population in the first excited subband and (c) plots are deep in the $\mathrm{N}_1$ phase.

The plots in Fig.~\ref{fig:susc-nn-ML} indicate that the density-wave modes with zero transverse momenta experience most enhancement from the attractive inter-layer interactions. This is an expected result given that density-wave fluctuations are in-plane density modulations and at $k_z=0$, they are aligned across the layers and thus, experience the maximum softening due to inter-layer attraction. We note that one expects the reverse scenario, i.e. maximum {\it suppression} of density-waves at $k_z=0$, had the inter-layer interactions been repulsive (such as multi-layer systems of 2DEG).

On the other hand, the odd ripplons are expected to experience most softening at $k_z d = \pi$ which corresponds to dimerization. At $k_z d = \pi$, the out-of-plane bumps of even numbered layers lie closest to the valleys of odd numbered layers, forming an energetically favorable configuration (shown schematically in Fig.~\ref{fig:phases}b). The slightly higher peak of 01-ripplon response function at $k_z d = \pi$ compared to $k_z = 0$ is noticeable in Fig.~\ref{fig:susc-01-ML}b-c. The smaller peak in the 01-ripplon response function (visible for $0 < k_z \lesssim \pi/2d$) is due to coupling to the softened density-wave mode at $q = 2\,k_{F,1}$.\\

In all of the studied cases, although the ripplon softening was found to be a more pronounced effect in multi-layer configurations compared to single-layer systems, the density-wave instability still precedes the ripplon instability. The first density-wave mode that becomes unstable has zero transverse momentum, implying that the density-wave and ripplon fluctuations are decoupled. Therefore, the density ordered phase to follow does not necessarily have out-of-plane order. In the remainder of this section, we discuss the phase diagrams of multi-layer systems for three inter-layer separations $\sqrt{n}d = 2,~1.5,~1.25$.\\

Fig.~\ref{fig:phase-diag-ML-2.0} shows the phase diagram of a multi-layer configuration with $\sqrt{n}d~=~2$ and $N_l = 50$. The dashed black line on the left plot indicates the $\mathrm{N}$-$\mathrm{DW}$ boundary line of the single-layer system (copied from Fig.~\ref{fig:phase-diag-SL}). As mentioned before, the first unstable mode is an in-plane density-wave mode with zero transverse momentum. We also find the most noticeable deviation of the $\mathrm{N}$-$\mathrm{DW}$ phase boundary occurs for larger values of $a_{\perp}$. The switching of the unstable wave-vector from $q=2k_{F,0}$ to $q=2k_{F,1}$ in the $\mathrm{N}_1$ phase is also found to occur for a smaller value of $\sqrt{n}a_\perp$ compared to the single-layer case, i.e. closer to the $\mathrm{N}_0$-$\mathrm{N}_1$ boundary.

Fig.~\ref{fig:phase-diag-ML-1.5} shows the phase diagram for $\sqrt{n}d~=~1.5$ and $N_l = 50$. The hatched region indicates the configurations at which the inter-layer tunneling is not negligible anymore and the approximation of independent layers is not justified. The physically interesting part of the phase diagram, however, lies outside of the hatched region. We notice that the $\mathrm{N}$$-$$\mathrm{DW}$ boundary line deviates even further from that of single-layer systems. The switching point of the unstable wave-vector lies very close to $\mathrm{N}_0$$-$$\mathrm{N}_1$ boundary. In other words, the $\mathrm{N}_1$ liquid phase always goes unstable due to the softened density-wave mode at $q=2k_{F,1}$. Since $k_{F,1} = 0$ along the $\mathrm{N}_0$$-$$\mathrm{N}_1$ transition line, the unstable wave-vector can be arbitrarily small in the vicinity of the switching point (see Fig.~\ref{fig:phase-diag-ML-1.5}b).

A more dramatic behavior is observed for smaller layer separations. Fig.~\ref{fig:phase-diag-ML-1.25} shows the phase diagram for $\sqrt{n}d~=~1.25$ and $N_l = 50$. It is noticed that the $\mathrm{N}$$-$$\mathrm{DW}$ boundary line becomes virtually tangent to the $\mathrm{N}_0$$-$$\mathrm{N}_1$ transition line in the range $0.21 < \sqrt{n}a_{\perp} < 0.26$. Along this part of the phase boundary, the transition to the inhomogeneous phase is driven by extremely long wavelength density-wave modes.\\

In the next section, we approach the same problem again using an approximate RPA-like formalism. Although we do not expect quantitatively reliable results, we still find that such an approach yields interesting analytical insights into some of the peculiar results of this section, in particular, the sudden switching of the unstable mode along the $\mathrm{N}$-$\mathrm{DW}$ boundary and the appearance of long wavelength unstable modes in multi-layer systems.

\section{Insights from the RPA approximation: neglecting short-range exchange interactions}\label{sec:RPA}
The major quantitative results of this paper was presented in the preceding section by numerically evaluating the response functions in the TDHF approximation. However, some of the results do not appeal to immediate intuition. In particular, (i) in single-layer systems, starting from the $\mathrm{N}_1$ phase, it is not clear why the the unstable density-wave abruptly switches from $q=2k_{F,0}$ to $q=2k_{F,1}$ as the population of particles in the first subband is increased (see Fig.~\ref{fig:phase-diag-SL}) and, (ii) the appearance of extremely long wavelength unstable density-wave modes along certain parts of the $\mathrm{N}$-$\mathrm{DW}$ phase boundary in multi-layer systems is puzzling. In this section, we develop a simplistic and minimal model by applying successive approximations to the TDHF formalism to derive an RPA-like expression for the density-density response function, using which we will qualitatively explain the above findings.

We start by noting that the main difficulty in obtaining analytical expressions in the TDHF approximation is the exact treatment of exchange interactions, i.e. solving the Bethe-Salpeter integral equation. In the RPA approximation, on the other hand, one completely neglects the exchange interactions and this difficulty does not arise. However, the RPA approximation is not readily applicable to our problem, given that large cancellations are expected between the direct and exchange interactions of particle-hole pairs. This can be easily seen in the simplest case, i.e. a single-layer system in the two-dimensional limit ($a_\perp \rightarrow 0$). In this limit, the only relevant interaction matrix element is $\mathcal{V}_{00;00}(\qq) = 4\sqrt{2\pi}D^{2}/3 a_{\perp} - 2\pi D^{2} q e^{-q|n|d} + \mathcal{O}(D^2q^2a_{\perp})$. The {\it s}-wave component of $\mathcal{V}_{00;00}(\qq)$ diverges in the limit $a_\perp \rightarrow 0$. In a system of spinless fermions (which is the case here), the {\it s}-wave interactions between the particles must vanish due to Fermi statistics and this cancellation only happens if one considers both direct and exchange interactions in a balanced way. This is clearly not the case in the RPA approximation. In these cases, it is customary to resort to heuristic methods to capture the exchange effects in an approximate way. Hubbard-type many-body local field approximations are widely used in the study of electron liquid~\cite{Mahan} and have also been generalized to quasi-two-dimensional systems~\cite{Yarlagadda1994}. Such approximations, however, essentially aim at improving the long wavelength behavior of the response functions. In our problem, we are interested in the response to density wave fluctuations at wavelengths in the order of the inverse Fermi momentum. Therefore, the many-body local fields used for electronic systems are not readily applicable to our problem and must be modified.  

\begin{figure}[t]
\includegraphics[width=7cm]{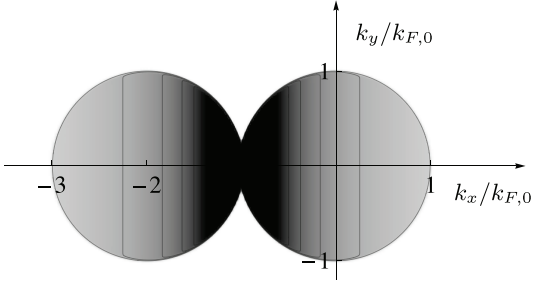}
\caption{Density plot of $\Pi^{(0)}_{00;00}(2k_{F,0}\hat{x},\mathbf{k})$ showing the singular behavior at $\kk=-k_{F,0}\hat{x}$.}
\label{fig:PIplot}
\end{figure}

Since we are only interested in qualitatively relevant results in this section, we take the easiet route and argue that by simply removing the {\it s}-wave component from all of the interaction matrix elements, the RPA formalism yields reasonably decent values for the density-density response function at $q \approx 2\,k_{F,j}$. This claim can be justified by investigating the Bethe-Salpeter equation for the irreducible polarization with more care. For the clarity of argument, we consider the single-subband limit first, where the bookkeeping of subband indices can be obviated. Taking the static limit, $\nu_n \rightarrow 0$, and defining $f(\qq,\kk) \equiv \Pi^\star_{00;00}(\qq,i0^+;\kk)/\Pi^{(0)}_{00;00}(\qq,i0^+;\kk)$, one can rewrite Eq.~(\ref{eq:integeq1}) as:
\begin{equation}\label{eq:integeqapprox}
f(\qq,\kk) =  1 - \int\frac{\mathrm{d}^2\kk'}{(2\pi)^2}\,u(\kk'-\kk)\,\Pi^{(0)}(\qq,\kk')\,f(\qq,\kk'),
\end{equation}
where $u(\kk) \equiv \mathcal{V}_{00;00}(\kk)$ and:
\begin{align}\label{eq:PI0}
\Pi^{(0)}(\qq,\kk) &\equiv \Pi^{(0)}_{00;00}(\qq,i0^+;\kk)\nonumber\\
&=\frac{n^{F}\big(\tilde{\xi}_{\kk,0}\big)-n^{F}\big(\tilde{\xi}_{\kk+\qq,0}\big)}{\tilde{\xi}_{\kk,0} - \tilde{\xi}_{\kk+\qq,0} + i0^+}.
\end{align}
For concreteness, we set $\qq = 2 k_F \hat{x}$. According to Eq.~(\ref{eq:PI0}) and as shown in Fig.~\ref{fig:PIplot}, $\Pi^{(0)}(2k_F\hat{x},\kk)$ is singular at $\kk_0=-k_F\hat{x}$ and we expect the most important contributions to the integral on the right hand side of Eq.~(\ref{eq:integeqapprox}) to result from the regions in the vicinity of $\kk_0$. Thus, we may approximately replace $u(\kk'-\kk)$ with $u(\kk'-\kk_0)$ in the integrand. On the other hand, $\Pi^\star_{00;00}(2k_F\hat{x},i0^+) = \int_{\kk'}f(2k_F\hat{x},\kk')\,\Pi^{(0)}(2k_F\hat{x},\kk')$ by definition, in which we may again approximately replace $f(2k_F\hat{x},\kk')$ with $f(2k_F\hat{x},\kk_0)$ according to same argument. Combining both approximations, we find that the final recipe is to replace $u(\kk'-\kk)$ with $u(0)$ in Eq.~(\ref{eq:integeqapprox}), i.e. to keep only the long-range exchange interactions. We emphasize that the above argument is special to the analysis of $|\qq| \approx 2 k_F$ modes. 

Once the short-range exchange interactions are neglected, the Bethe-Salpeter equation can be trivially solved. Combining this results with the Dyson's equation, Eq.~(\ref{eq:dysonlayer}), we find that the only effect of the long-range exchange interactions is to remove the {\it s}-wave component from the interaction matrix elements, as we expected.

The above argument can be easily generalized to multi-subband and multi-layer systems using a matrix notation and we omit it here. In brief, we find that the general recipe is to simply make the substitution $\mathcal{V}_{\rho\nu;\mu\sigma}(\kk-\kk') \rightarrow \mathcal{V}_{\rho\nu;\mu\sigma}(\mathbf{0})$ in the Bethe-Salpeter equation, yielding the following linear algebraic system of equations:
\begin{equation}\label{eq:xch}
\Pi^{\star} = \Pi^{(0)} - \Pi^{(0)}\mathcal{V}_{\mathrm{xch}}\Pi^{\star}.
\end{equation}
We have dropped the common arguments and subband indices for brevity in the above equation. Also, matrix multiplication is implied in each pair of subband indices. The approximate exchange interaction matrix, $\mathcal{V}_{\mathrm{xch}}$, is defined as:
\begin{equation}
\left[\mathcal{V}_{\mathrm{xch}}\right]_{\mu\nu;\rho\sigma} = \mathcal{V}_{\rho\nu;\mu\sigma}(\mathbf{0}).
\end{equation}
Combining Eqs.~(\ref{eq:xch}) and (\ref{eq:dysonlayer3}), we get:
\begin{equation}\label{eq:piapproxfinal}
\tilde{\Pi}^{(k_z)} = \tilde{\Pi}^{(0)} + \tilde{\Pi}^{(0)}\big(\tilde{\mathcal{V}}^{(k_z)} - \mathcal{V}_{\mathrm{xch}}\big)\tilde{\Pi}^{(k_z)}.
\end{equation} 
To ensure no violation of conservation laws, the short-range exchange interactions must also be neglected in the self-energy corrections. However, the long-range direct and exchange intra-layer exchange cancel each other. As mentioned in Sec.~\ref{sec:HFML}, the direct inter-layer interactions merely shift the zero-point subband energies by a small amount and for simplicity, one may neglect such corrections as well. Therefore, self-energy corrections can be neglected altogether. We refer to this approximation scheme as \RPA for brevity, with the $\mathrm{ns}$ subscript indicating the absence of {\it s}-wave interaction terms.\\

The important features of the phase diagrams presented in the previous section can be captured by keeping only the first two subbands. We also restrict the forthcoming analysis to $k_z=0$, given that such modes become unstable first. Under such assumptions, Eqs.~(\ref{eq:piapproxfinal}) and~(\ref{eq:chiddpolk}) yield:
\begin{multline}\label{eq:RPAdd}
\chi_{\mathrm{dd}}(q)=\frac{-1}{\mathrm{det}_{\Pi}(q)}\Big[\Pi^{(0)}_{00}(q)+\Pi^{(0)}_{11}(q)+\Pi^{(0)}_{00}(q)\Pi^{(0)}_{11}(q)\\
\times\big(\mathcal{V}_{00;00}^{\mathrm{eff}}(q)+\mathcal{V}_{11;11}^{\mathrm{eff}}(q)-2\mathcal{V}_{00;11}^{\mathrm{eff}}(q)\big)\Big],
\end{multline}
where:
\begin{multline}\label{eq:det}
\mathrm{det}_{\Pi}(q) = 1 - \mathcal{V}_{00;00}^{\mathrm{eff}}(q)\Pi^{(0)}_{00}(q) - \mathcal{V}_{11;11}^{\mathrm{eff}}(q)\Pi^{(0)}_{11}(q)\\
+ \Pi^{(0)}_{00}(q) \Pi^{(0)}_{11}(q)\left(\mathcal{V}_{00;00}^{\mathrm{eff}}(q)\mathcal{V}_{11;11}^{\mathrm{eff}}(q)-\mathcal{V}_{00;11}^{\mathrm{eff}}(q)^2\right).
\end{multline}
The effective interaction matrix elements, $\mathcal{V}_{\alpha\beta;\gamma\lambda}^{\mathrm{eff}}$, are defined as:
\begin{equation}\label{eq:Veff}
\mathcal{V}_{\alpha\beta;\gamma\lambda}^{\mathrm{eff}}(q) = \left[\sum_{n=-N_l+1}^{N_l-1}\mathcal{V}^{(n)}_{\alpha\beta;\gamma\lambda}(q)\right] - \mathcal{V}_{\gamma\beta;\alpha\lambda}(0),
\end{equation}
and the bare static intra-subband polarization, $\Pi^{(0)}_{\alpha\alpha}(\qq)$, can be evaluated analytically in the absence of self-energy corrections:
\begin{eqnarray}\label{eq:chi0}
\Pi^{(0)}_{\alpha\alpha}(\qq) & = & \int \frac{\mathrm{d}^2\kk}{(2\pi)^2}\frac{n^F(\xi^0_{\kk+\qq,\alpha})-n^F(\xi^0_{\kk,\alpha})}{\xi^0_{\kk+\qq,\alpha} - \xi^0_{\kk,\alpha} + i0^+}\nonumber\\
&=&\frac{m}{2\pi\hbar^2}\left(1-\sqrt{1-\left(\frac{2k^{(0)}_{F,\alpha}}{q}\right)^2}\theta(q-2k^{(0)}_{F,\alpha})\right).\nonumber\\
\end{eqnarray}
In the above equation, $\{k^{(0)}_{F,\alpha}\}$ are the Fermi momenta of a non-interacting quasi-two-dimensional gas, as shown in Table~\ref{tab:kf}.

\begin{figure}[t!]
\includegraphics[width=7cm]{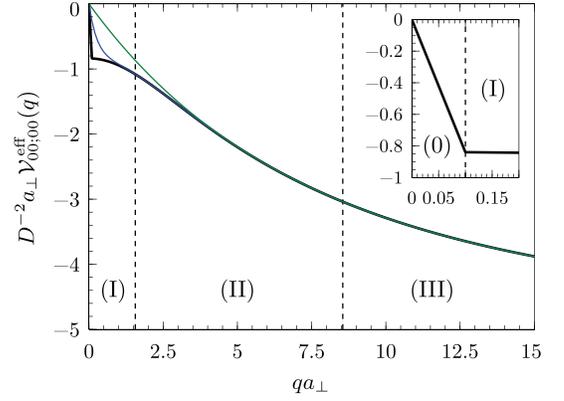}
\caption{The effective interaction matrix element $\mathcal{V}_{00;00}^{\mathrm{eff}}(q)$ vs. $q$. The green (upper), blue (middle) and black (lower) lines correspond to $N_l=1,~3$ and $200$ respectively. In all cases, $a_{\perp}/d = 1/15$. The inset plot shows the $N_l=200$ case at small values of $q$ for clarity. In the IL limit ($a_\perp \ll d$), one can classify the length scales into four categories according to the behavior of effective interactions, as indicated on the figure.  
{\it Category~(0):} [$q \lesssim L^{-1}$] length scales larger than $L$. The layered structure of the stack is invisible to density-wave fluctuations in this length scale. Since we have set $k_z=0$, the in-plane density waves are all aligned across the layers and collectively behave like a single density-wave with an effective dipolar interaction strength of $(2N_l-1) D^2$ (Ref. to Eq.~\ref{eq:veffcond2}). In other words, the whole stack behaves like a single two-dimensional layer;
{\it Category~(I):} [$L^{-1} \lesssim q \lesssim d^{-1}$] length scales smaller than $L$ and larger than inter-layer separation $d$. Density wave fluctuations in any given layer interacts with a fraction $(qL)^{-1}$ of other layers, hence, resulting in a constant scale invariant effective interaction $-2\pi D^2 (2N_l - 1) q \times (qL)^{-1} \approx -4 \pi D^2/d$;
{\it Category~(II):} [$d^{-1} \lesssim q \lesssim a_{\perp}^{-1}$] length scales smaller than $d$ and larger than $a_\perp$. In this regime, the inter-layer interactions are exponentially small (see Eq.~\ref{eq:Vexp}) and the density waves only interact within the layers.
{\it Category~(III):} [$q \lesssim a_{\perp}^{-1}$] length scales smaller than $a_\perp$. Each of the interaction matrix elements ($\mathcal{V}_{00;00}^{\mathrm{eff}}$, $\mathcal{V}_{00;11}^{\mathrm{eff}}$, etc.) assume different non-universal constant values in the order of $D^2/a_\perp$.}
\label{fig:veff}
\end{figure}

\begin{table}[h!]
\begin{tabular}{c|c|c}
& $a_{\perp} < \frac{1}{\sqrt{2 \pi n}}$ & $\frac{1}{\sqrt{2 \pi n}} \leq a_{\perp} < \frac{\sqrt{3}}{\sqrt{2 \pi n}} $\\
\hline
$k^{(0)}_{F,0}$ & $\sqrt{4\pi n}$ & $\sqrt{2\pi n + n/a_{\perp}^2}$\\
\hline
$k^{(0)}_{F,1}$ & 0 & $\sqrt{2\pi n - n/a_{\perp}^2}$
\end{tabular}
\caption{The Fermi momenta of the first two subbands of a non-interacting quasi-two-dimensional gas.}
\label{tab:kf}
\end{table}

In this simplified approach, the single-layer and multi-layer systems are treated likewise. The multi-layer effects are included in the effective interactions. In other words, $\mathcal{V}^\mathrm{eff}_{\alpha\beta;\gamma\lambda}$ is the sum of all intra-layer and inter-layer interactions. Studying the behavior of the effective interactions is thus a key step in understanding the difference between the phase diagrams of single-layer and multi-layer systems. We focus on the behavior of $\mathcal{V}_{00;00}^{\mathrm{eff}}(q)$, which is find to be qualitatively identical to the behavior of the rest of the involved interaction matrix elements, $\mathcal{V}_{11;11}^{\mathrm{eff}}(q)$ and $\mathcal{V}_{00;11}^{\mathrm{eff}}(q)$. Expanding Eq.~(A4) about $q=0$, we get:
\begin{equation}\label{eq:Vexp}
\mathcal{V}_{00;00}^{(n)}(q) = \frac{4\sqrt{2\pi}D^{2}}{3 a_{\perp}}\delta_{n,0} - 2\pi D^{2} q e^{-q|n|d} + \mathcal{O}(q^2 a_{\perp}),
\end{equation}
using which we find:
\begin{equation}\label{eq:veffcond}
\mathcal{V}_{00;00}^{\mathrm{eff}}(q) \simeq \left\{\begin{tabular}{ll}
$\displaystyle-2\pi D^{2} q $ & \quad $N_l=1$\\
$\displaystyle-2\pi D^{2} q\,\coth\left(\frac{qd}{2}\right)$ & \quad $N_l =\infty,$\end{tabular}\right.
\end{equation}
where the neglected terms are $\mathcal{O}(q^2 a_{\perp})$. For future reference, it is also useful to study the behavior of $\mathcal{V}_{00;00}^{\mathrm{eff}}(q)$ for finite $N_l$, and for wavelengths longer than inter-layer separation $d$. In this limit, we find:
\begin{equation}\label{eq:veffcond2}
\mathcal{V}_{00;00}^{\mathrm{eff}}(q) \simeq \left\{\begin{tabular}{ll}
$\displaystyle-2\pi D^{2} (2N_l-1) q $ & \quad $q \lesssim L^{-1},$\\
$\displaystyle-\frac{4\pi D^{2}}{d}$ & \quad $L^{-1} \lesssim q \lesssim d^{-1}.$
\end{tabular}\right.
\end{equation}
Again, the neglected terms are $\mathcal{O}(q^2 a_{\perp})$. We remind that $L \equiv N_l d$ is the transverse size of the stack. While the effective interaction in single-layer systems has a linear dependence on $q$ in the regime $q \ll a_{\perp}^{-1}$, its behavior is very different for long wavelength modes in multi-layer systems. Fig.~\ref{fig:veff} shows $\mathcal{V}_{00;00}^{\mathrm{eff}}(q)$ as a function of $q$ for three different number of layers, $N_l=1$ (green), 3 (blue) and 200 (black). In the IL limit ($a_\perp \ll d$), one can classify the length scales into four regimes according to the behavior of effective interactions. These regimes are indicated in Fig.~(\ref{fig:veff}) and a brief description for each is provided in the caption. Consequently, one can categorize the density wave fluctuations according to the same length scale classification and as we will see shortly, this is a key step in interpreting the features of the obtained phase diagrams.\\

We start the analysis of \RPA with the simpler case of single-layer systems. The stability of the normal phase can be determined by looking at the behavior of $\mathrm{det}_{\Pi}(q)$, which is the term appearing in the denominator of the \RPA expression for the density-density response function. For small interaction strengths, $\mathrm{det}_{\Pi}(q) \approx 1$. Upon increasing the interactions, $\mathrm{det}_{\Pi}(q)$ decreases and eventually, crosses zero at some $q$, signaling the appearance of a softened mode.

Generally, we found that the approximate identity $\mathcal{V}_{00;00}^{\mathrm{eff}}(q)\mathcal{V}_{11;11}^{\mathrm{eff}}(q)~\approx~\mathcal{V}_{00;11}^{\mathrm{eff}}(q)^2$ holds well for all $q$. In particular, all interaction matrix elements behave similarly in the limit $q \lesssim a_\perp^{-1}$ according to the remarks given in the caption of Fig.~\ref{fig:veff}, justifying this identity for wavelengths longer than $a_\perp$. The second line of Eq.~(\ref{eq:det}) can be neglected in light of the this observation, yielding the following simple expression for $\mathrm{det}_{\Pi}(q)$:
\begin{equation}\label{eq:detapprox}
\mathrm{det}_{\Pi}(q) \approx 1 - \mathcal{V}^{\mathrm{eff}}_{00;00}(q)\Pi^{(0)}_{00}(q) - \mathcal{V}^{\mathrm{eff}}_{11;11}(q)\Pi^{(0)}_{11}(q),
\end{equation}
Intuitively, the above equation implies that the net density-wave enhancement is the algebraic sum of RPA-like density-wave enhancement of each subband. 

\begin{figure}[h!]
\includegraphics[width=7cm]{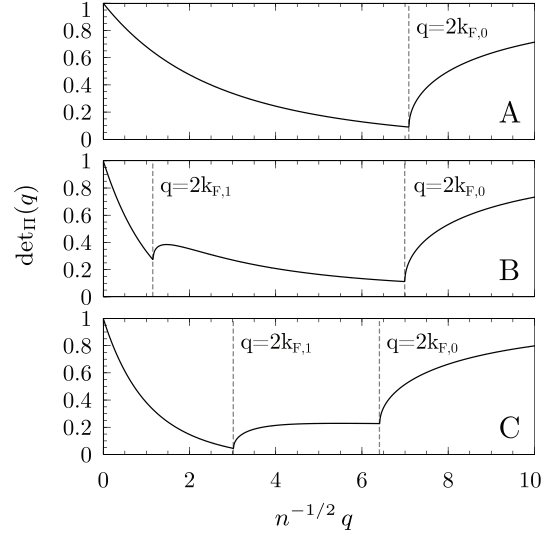}
\caption{Plot of $\mathrm{det}_{\Pi}(q)$ vs. $n^{-1/2}q$ for three points in the normal phase (Ref. to Eq.~\ref{eq:RPAdd} and~\ref{eq:det} and the following text for details): (A) $\mathcal{N}_0$ phase [$r_d=0.40$, $\sqrt{n}a_{\perp}=0.30$] (B) $\mathcal{N}_1$ phase [$r_d=0.50$, $\sqrt{n}a_{\perp}=0.41$] (C) $\mathcal{N}_1$ phase [$r_d=0.50$, $\sqrt{n}a_{\perp}=0.50$]. Refer to Fig.~\ref{fig:RPA-phase-diag-SL} to see the location of these three points in the phase diagram.}
\label{fig:det-PI}
\end{figure}

Fig.~\ref{fig:det-PI} shows the plot of $\mathrm{det}_{\Pi}(q)$ for a point in $\mathrm{N}_0$ phase (A) and two points in $\mathrm{N}_1$ phase, with small and large population of the first subband (B and C respectively). In it noticed that in (A) and (B), the most softened mode (i.e., smaller $\mathrm{det}_{\Pi}$) is $q=2k_{F,0}$, while in (C), $q=2k_{F,1}$ is the most softened. The shift of the unstable mode from $q=2k_{F,0}$ to $q=2k_{F,1}$ in the $\mathrm{N}_1$ phase can be explained in light of Eqs.~(\ref{eq:detapprox}) and Eq.~(\ref{eq:chi0}). In order to simply the discussion, we note that as long as $q<2k_{F,\alpha}$, $\Pi^{(0)}_{\alpha\alpha}(q)$ has a positive constant value and rapidly falls for $q$ larger than $2k_{F,\alpha}$. Thus, one only needs to monitor $\mathrm{det}_{\Pi}(q)$ for $q=2k_{F,0}$ and $q=2k_{F,1}$, where the product of the effective interactions and the bare polarizations is be the largest. There are two possible scenarios in the $\mathrm{N}_1$ phase:

{\em Case I} ($k_{F,1} \ll k_{F,0}$): this case corresponds to dilute quasiparticles in the first excited subband and consequently, the effective interactions (which increase linearly with momentum) are weak at $q=2k_{F,1}$. Therefore, the sum of RPA-like enhancements resulting from both subbands at $q=2k_{F,1}$ is smaller than the enhancement resulting mainly from the zeroth subband at $q=2k_{F,0}$ (see Fig.~\ref{fig:det-PI}B). Since $k_{F,1} \ll k_{F,0}$, $\Pi^{(0)}_{11}(2k_{F,0}) \approx 0$ and at $q=2k_{F,0}$, the density-wave enhancements are mainly due to the interactions in the zeroth subband.

{\em Case II} ($k_{F,1} \sim k_{F,0}$): this situation arises when there is a significant population in the first excited subband, i.e. deep in the $\mathrm{N}_1$ phase. The scenario is reversed in this case and the sum of enhancements resulting from both subbands at $q=2k_{F,1}$ is larger than the enhancement resulting mainly from the zeroth subband at $q=2k_{F,0}$ (see Fig.~\ref{fig:det-PI}C).

It is not hard to see that the second scenario may only happen if the rise of interactions is {\em slower} than the fall of density of particle-hole excitations as a function of $q$. The linear momentum dependence of dipolar interactions and the rapid fall of $\Pi^{(0)}_{11}(q)$ for $q>2k_{F,1}$ guarantees the realization of this situation for large enough values of $k_{F,1}$.\\

\begin{figure}[b!]
\includegraphics[width=7cm]{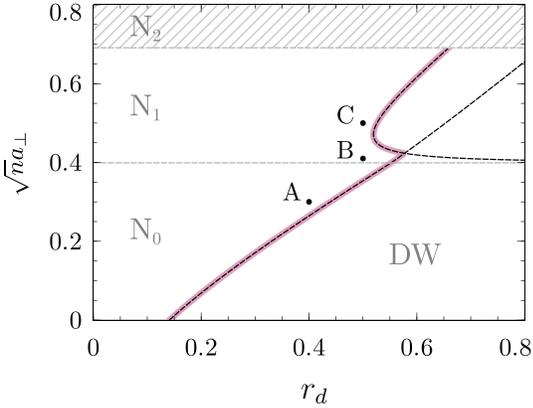}
\caption{The approximate phase diagram of quasi-two-dimensional dipolar fermions in a single-layer configuration in the \RPA approximation. The dashed lines show the smallest $r_d$ for which the density-wave mode at $q=2k_{F,0}$ or $q=2k_{F,1}$ become unstable. The pink line indicate the first unstable mode. The switching of unstable density-wave mode in the $\mathrm{N}_1$ phase is noticeable. Refer to Fig.~\ref{fig:det-PI} for a plot of $\mathrm{det}_{\Pi}(q)$ for the three points marked in the diagram.}
\label{fig:RPA-phase-diag-SL}
\end{figure}

Fig.~\ref{fig:RPA-phase-diag-SL} shows the approximate phase diagram of a single-layer system calculated in the \RPA approximation. The flatness of $\mathrm{N}_0$-$\mathrm{N}_1$ and $\mathrm{N}_1$-$\mathrm{N}_2$ boundaries is due to ignoring the self-energy corrections in the normal phase. There is a striking similarity between this phase diagram and the one obtained by exact numerical calculation of TDHF response functions (Fig.~\ref{fig:phase-diag-SL}). However, the predicted value for the DW instability at $a_\perp \rightarrow 0$, $r_d^{\mathrm{RPA}} \approx 0.15$, is more than a factor of two smaller than the same value predicted within TDHF, $r_d^{\mathrm{TDHF}} \approx 0.39$.\\

\begin{figure}[h!]
\includegraphics[width=7cm]{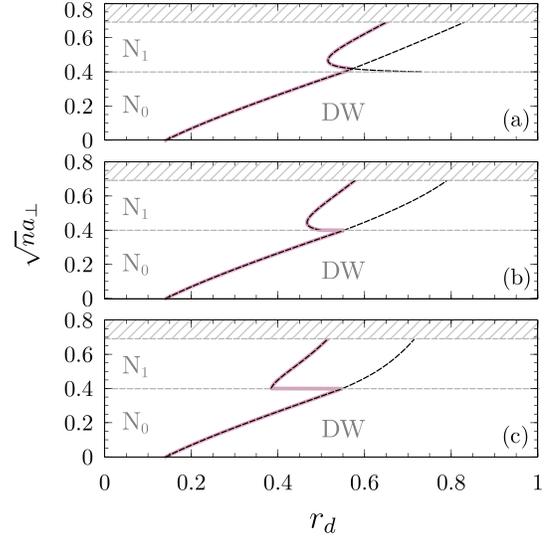}
\caption{The approximate phase diagram of quasi-two-dimensional dipolar fermions in multi-layer configurations ($N_l=20$) in the \RPA approximation: (a) $\sqrt{n}d = 3$ (b) $\sqrt{n}d = 2$ (c) $\sqrt{n}d = 1.5$. The dashed lines show the smallest $r_d$ for which the density-wave mode at $q=2k_{F,0}$ or $q=2k_{F,1}$ become unstable. The pink lines indicate the first unstable mode.}
\label{fig:RPA-phase-diag-ML}
\end{figure}

Fig.~\ref{fig:RPA-phase-diag-ML} shows the approximate phase diagram of three multi-layer systems with different inter-layer separations obtained using the \RPA scheme. It is noticed that the non-trivial features of multi-layer phase diagrams, i.e. (1) indifference of $\mathrm{N}$-$\mathrm{DW}$ boundary line to existence of multiple layers deep in the $\mathrm{N}_0$ phase, and (2) enhancement of density wave instability along parts of $\mathrm{N}_0$-$\mathrm{N}_1$ phase for smaller inter-layer separations, are also present in the picture that \RPA suggests.

The former feature can be explained by first noting that the studied range of inter-layer separations is such that $\sqrt{n}d \sim \mathcal{O}(1)$. Therefore, the unstable vector in the $\mathrm{N}_0$ phase, $q=2k_{F,0} \sim 2\sqrt{4 \pi n}$, is almost an order of magnitude larger than $d^{-1}$ and belong to category (II). The inter-layer interactions are irrelevant in this regime and the physics is identical to that of a single-layer system.

As a side note, we remark that in order to see the effect of inter-layer interaction in the single-subband limit ($\mathrm{N}_0$ phase), one must choose $d$ such that $\sqrt{n} d \ll 1$. In particular, in the limit $\sqrt{n} d \ll N_l^{-1}$, the unstable modes will lie in category (0) and the density wave instability will be driven by fluctuations whose length scale is larger than the transverse length of the stack. As mentioned earlier, the effective interactions are enhanced proportionally to the number of layers in this limit and as a result, we expect the interaction strength required for the onset of density-wave instability to be reduced by a factor of $\sim N_l^{-1}$.
   
The latter feature, i.e. appearance of long wavelength unstable modes close to $\mathrm{N}_0$-$\mathrm{N}_1$ boundary can be explained as follows. We discuss the simpler case of $N_l \rightarrow \infty$ first, in which all $q \lesssim d^{-1}$ lie inside category (I), i.e. where the effective interactions assume a constant value of $-4\pi D^2/d$. Existence of a small particle density $n_1$ in the first excited subband will result in the appearance of long wavelength gapless particle-hole excitations. The length scale associated to these modes can be very large and may as well lie within category (I) for small enough $n_1$, i.e. $q = 2k_{F,1} \sim 2\sqrt{4\pi n_1} \lesssim d^{-1}$. Since the density of long wavelength excitations is finite in two dimensions, i.e. $\lim_{|\qq|\rightarrow 0}\Pi^{(0)}(\qq) \sim \mathcal{O}(m/2\pi\hbar^2)$, they will have a finite RPA-like contribution of $\Pi^{(0)}_{11}(2k_{F,1}) \times \mathcal{V}^{\mathrm{eff}}_{11;11}(2k_{F,1}) \sim -2m D^2 / d \hbar^2$ to $\mathrm{det}_{\Pi}$ (see Eq.~\ref{eq:detapprox}). For small inter-layer separations, this contribution can be large and result in density-wave instability. 

In the limit $N_l \rightarrow \infty$, these modes appear exactly along the $\mathrm{N}_0$-$\mathrm{N}_1$ boundary where $q = 2k_{F,1} =0 $. The largest layer separation, $d_{\mathrm{max}}$, for which such long wavelength modes appear can be easily determined. At $d=d_{\mathrm{max}}$, the $q=0$ unstable mode appears only at one point, viz. at the intersection of $\mathrm{N}_0$-$\mathrm{N}_1$ and $\mathrm{N}$-$\mathrm{DW}$ lines. Therefore, both $q=2k_{F,0}$ and $q=0$ are unstable at this point (see Fig.~\ref{fig:RPA-phase-diag-ML}a). The \RPA instability condition at $q=0$ yields:
\begin{align}\label{eq:det0}
\mathrm{det}_{\Pi}(0^+) &= 1 - \frac{4\pi D^2}{d_{\mathrm{max}}}\left(\Pi^{(0)}_{00}(0)+\Pi^{(0)}_{11}(0)\right)\nonumber\\
&= 1 - \frac{4\pi D^2}{d_{\mathrm{max}}}\frac{m}{\pi \hbar^2}\nonumber\\
&= 0,
\end{align}
and the instability of $q = 2k_{F,0} = 2\sqrt{4\pi n}$ yields:
\begin{align}\label{eq:det1}
\mathrm{det}_{\Pi}(2k_{F,0}) &= 1 - \mathcal{V}^{\mathrm{eff}}_{00;00}(2\sqrt{4\pi n})\,\frac{m}{2\pi\hbar^2}\nonumber\\
&= 0.
\end{align}
In the above equation, the effective interaction must be evaluated on the $\mathrm{N}_0$-$\mathrm{N}_1$ boundary, i.e. $a_{\perp} = 1/\sqrt{2\pi n}$, i.e.  The simultaneous solution of these equations yields:
\begin{equation}\label{eq:solcrit}
\sqrt{n}d_{\mathrm{max}} \simeq 2.209, \quad r_d \simeq 0.5523.
\end{equation}
For $d<d_{\mathrm{max}}$, the $q=0$ unstable modes appear along a finite interval on the $\mathrm{N}_0$-$\mathrm{N}_1$ boundary (see Fig.~\ref{fig:phase-diag-ML-1.25} and Fig.~\ref{fig:RPA-phase-diag-ML}b-c).

The prediction of $d_{\mathrm{max}}$ within \RPA is significantly larger than the one inferred from TDHF calculations presented earlier (see Fig.~\ref{fig:phase-diag-ML-1.5}; $\sqrt{n}d_{\mathrm{max}}^{\mathrm{TDHF}} \simeq 1.5$ and $r_d \simeq 1.35$). This deviation is again due to approximate treatment of exchange interactions.\\

We conclude this section by briefly studying the scaling dependence of the wavevector of the long wavelength unstable modes discussed above on $N_l$. For finite $N_l$, $q=2k_{F,1}$ mode lies inside category (0) if the first excited subband is dilute enough. In this limit, the whole stack behaves collectively like a single pancake, with an effective interaction of $\sim - 2\pi (2N_l-1) D^2 q$. Assuming $q<L^{-1}$, the \RPA instability condition yields: 
\begin{eqnarray}
\mathrm{det}_{\Pi}(q) &\approx& 1 - 2\times\frac{m}{2\pi\hbar^2}\,2\pi D^2 (2N_l-1) q = 0.
\end{eqnarray}
Solving for $q$, we get:
\begin{equation}\label{eq:qcond}
q \approx \frac{\sqrt{n}}{2(2N_l-1)r_d}, \quad \frac{\sqrt{n}d}{4} \lesssim r_d \lesssim 0.55.
\end{equation}
The constraints imposed on $r_d$ in the above equation result from two requirements: on one hand, the solution must satisfy $q \lesssim L^{-1}$. On the other hand, the $r_d$ required for instability of this mode must be smaller than that required for the instability of the $q=2k_{F,0}$ mode, which is $\approx 0.55$ in the vicinity of the $\mathrm{N}_0$-$\mathrm{N}_1$ boundary and for $d$ not much less than $d_{\mathrm{max}}$ (see Eq.~\ref{eq:solcrit} and Fig.~\ref{fig:RPA-phase-diag-ML}). In the limit $N_l \rightarrow \infty$, the unstable wave-vector becomes 0.

\section{Discussion and Conclusions}\label{sec:disc}
In this paper, we studied the mean-field density ordering instabilities of quasi-two-dimensional fermionic polar molecules in both single-layer and multi-layer configurations. The dipole moments of the molecules were assumed to be aligned perpendicular to the confining planes using a dc electric field. We located the instabilities by evaluating various linear response functions in the liquid phase and searching for the softened modes. We considered both in-plane and out-of-plane density ordering instabilities, as schematically depicted in Fig.~\ref{fig:phases}.\\

In all of the studied cases, the instability of the in-plane density wave modes was found to precede the instability of out-of-plane ``ripplon'' modes, although the latter modes were also softened to some degree. We also found that leading unstable mode in multi-layer systems has a zero transverse momentum, i.e. the in-plane density-waves are aligned across the layers.\\

In multi-layer configurations, an interesting finding was the enhanced density wave instability driven by dilute quasiparticles of the first excited subband. By analyzing the effective interactions at various length scales in Sec.~\ref{sec:RPA}, we found that these dynamical instabilities are associated to the softening of low-energy particle-hole excitations whose wavelength is comparable to or larger than the transverse size of the system, $L$. On one hand, the density of such excitations is finite due to the underlying two-dimensionality of the system. On the other hand, their effective interaction is enhanced proportionally to the number of layers due to their long wavelength. Hence, they produce a significant density-wave softening effect.\\

Another interesting feature of the phase diagram of both single-layer and multi-layer configurations is the non-monotonicity of the $\mathrm{N}$-$\mathrm{DW}$ phase boundary as a function of transverse confinement width (see Figs.~\ref{fig:phase-diag-SL},~\ref{fig:phase-diag-ML-2.0}-\ref{fig:phase-diag-ML-1.25}). A phase diagram with similar qualitative features had been predicted before for quasi-2DEG using density functional theory with Perdew-Zunger type exchange-correlation energy~\cite{Ito2004}. Thus, we expect this feature of the presented instability diagram to persist in the true phase diagram, i.e. when the correlation effects are also taken into account. The protrusion of the liquid phase inside the density ordered phases allows realization of the following interesting experimental scenario: starting from a point such as $r_d=1.45,~\sqrt{n}a_\perp=0.15$ in a density ordered phase of a single-layer system, the homogeneous liquid phase can be revived upon relaxing the trap. However, the liquid state becomes unstable again upon relaxing the trap further (i.e. by crossing the red boundary line in Fig.~\ref{fig:phase-diag-SL}) toward a different density ordered phase.\\

As mentioned in the introduction, Yamaguchi {\it et al.}~\cite{Yamaguchi2010} and Sun {\it et al.}~\cite{Sun2010} have recently studied the density-wave instability of a (strictly) two-dimensional system of polar molecules ($a_\perp \rightarrow 0$) within the RPA approximation. Their study, however, includes the more general case of tilted dipoles and finite temperatures. At zero temperature and dipole tilt angle, their results indicated that the density-wave instability occurs at $r_d \approx 0.17$. The RPA-like approximation used in Sec.~\ref{sec:RPA} yields $r_d \approx 0.15$ in the limit $a_\perp \rightarrow 0$ which is in good agreement with the result of Ref.~\cite{Yamaguchi2010}.

The TDHF results presented in Sec.~\ref{sec:res}, however, predicts $r_d \approx 0.39$ in the same limit which is more than a factor of two larger than the prediction of the RPA approximation. Hence, we conclude that an exact treatment of short-range exchange interactions is important for quantitatively reliable predictions of phase transitions in dipolar systems.\\

The instabilities predicted in the mean-field picture must be interpreted with care. On one hand, one must consider the possibility that the actual phase transition is first-order. In this case, the mode softening criterion does not indicate the true transition but signifies the spinodal line (i.e. the end of liquid metastability region) and the actual phase transition occurs for smaller values of $r_d$. Typically, the spinodal line lies close to the actual transition line~\cite{Chaikin2000} and therefore, we do not expect the above issue to be a major source of error in the obtained phase diagrams.

The main shortcoming of the present analysis lies in the mean-field approximation and absence of correlation effects in the liquid phase. It is known that the mean-field mode softening analysis often predicts the transition to the symmetry broken phases too early. For instance, the Wigner crystal phase of 2D electrons with $1/r$ Coulomb interactions is found to become stable for $r_s \geq 1.44$ in the mean-field approximation~\cite{Trail2003} while a more realistic quantum Monte-Carlo calculation yields a value of $r_s \gtrsim 38$~\cite{Tanatar1989}. Thus, we expect that the instability lines shown in Figs.~\ref{fig:phase-diag-SL} and \ref{fig:phase-diag-ML-2.0}-\ref{fig:phase-diag-ML-1.25} will be shifted to larger values of $r_d$ once correlation effects are taken into account. Since mean-field predictions improve by increasing the dimensionality, this correction is expected to be smaller in multi-layer systems compared to single-layer systems. Nevertheless, we expect that the mean-field transition lines obtained here will describe sharp crossovers to the regime of strong short range crystal correlations (with no long-range order) in real systems, with the actual phase transitions following for larger values of $r_d$.

We remark that the regime of strong short-range crystal correlations with no long-range order is reminiscent of the pseudogap phase of fermions with strong attractive interactions. In the latter case, one finds short-range pairing correlations but no true long-range order, i.e. no condensation of molecules.

While the full analysis of such strongly correlated ``preformed density-wave'' state is outside the scope of this paper, the mean-field analysis presented here is an indispensable first step toward the study of this intriguing state.\\

At the time this paper is being written, the experimental verification of the presented results can still be challenging. The maximum dipolar interaction strength accessible in the experiments is $r_d \approx 0.05$, which belongs the experiments of the group at JILA with KRb polar molecules. Observation of density-wave instability either requires production of denser gases or using molecules with larger dipole moments (the permanent dipole moment of KRb is $0.55~D$). Further progress in experiments with LiCs~\cite{Deiglmayr2008} and RbCs~\cite{Sage2005} polar molecules whose permanent dipole moments are $5.5~D$ and $1.25~D$ respectively, are among the other promising experimental directions toward observation of the effects discussed in this paper.\\

We remark that in the same way ultracold atoms were utilized as a simulator for confined electrons with effective short-range interactions and shed light on the Hubbard model, ultracold polar molecules may be utilized as a tool to address unsettled questions regarding the nature of transitions to density ordered phases, intermediate strongly correlated states (such as the electron nematic phase) and microemulsion phases (such as stripes and bubbles)~\cite{Spivak2004}.\\

Throughout this work, we had assumed that the stable phase in the weakly interacting regime is the normal liquid phase. Neglecting the weak high angular momentum superconducting phases which may only appear at very low temperatures, this assumption is valid for single-layer systems. In multi-layer configurations, however, the normal phase is known to be unstable toward dimerized pseudo-gap and inter-layer superfluid phases~\cite{Potter2010,Pikovski2010,Baranov2011} at low temperatures. The passage through these phases occur through Ising-like, and Berezinskii-Kosterlitz-Thouless phase transitions, respectively. In our study, the configuration which is most likely to be in a superfluid phase in all of the presented multi-layer phase diagrams (Figs.~\ref{fig:phase-diag-ML-2.0}-\ref{fig:phase-diag-ML-1.25}; excluding the hatched regions) is $\sqrt{n}d = 1.25$, $r_d \approx 0.8$ and $\sqrt{n}a_\perp \approx 0.28$. For such a configuration, the critical temperature for BCS transition is estimated to be $T_c/T_{F}^{(0)} \approx 0.013$~\cite{BCSestimate} using the results of Ref.~\cite{Potter2010}. Therefore, the temperature chosen in this study, $T/T_F^{(0)} = 0.02$, is above the inter-layer pairing transition and our assumption about the stability of the liquid phase for weak interactions is justified.\\

The competition between inter-layer pairing instability and density-ordering instabilities at lower temperatures, or for systems with smaller inter-layer separations, is an interesting topic for future works. The results presented here can also be easily extended to tilted dipoles. Reducing the intra-subband repulsion and enhancing the inter-subband repulsion, tilting the dipoles may reverse the order of density-wave and ripplon instabilities. The competition between intra-layer {\it p}-wave superfluidity which is expected to occur for large tilt angles~\cite{Bruun2008}, ripplon and density-wave instabilities is another interesting topic for future research.\\

{\it Note added:} After the completion of this work, we became aware of a recent related work by Zinner {\it et al.}~\cite{Zinner2011} in which they study the density-wave instability of stacks of strictly two-dimensional polar molecules ($a_\perp \rightarrow 0$) within the RPA approximation.

\section{Acknowledgements}
The authors would like to thank Bertrand Halperin, Daw-Wei Wang, Julia Meyer and Sankar Das Sarma for insightful discussions. This work was supported by the Army Research Office with funding from the DARPA OLE program, Harvard-MIT CUA, NSF Grant No. DMR-07-05472, AFOSR Quantum Simulation MURI, AFOSR MURI on Ultracold Molecules and the ARO-MURI on Atomtronics.

\appendix

\section{Analytical expressions for the interaction matrix elements}\label{sec:app1}
In this appendix, we provide analytical expressions for inter-layer interaction between quasiparticles in the first two subbands. The interaction between quasiparticles in higher subbands may also be calculated using the same method.

Using Eq.~(\ref{eq:V}) and approximating the Wannier's wavefunctions by shifted harmonic oscillator wavefunctions, the effective interaction between particles confined to planes $z=0$ and $z=l$ can be expressed as:
\begin{eqnarray}\label{eq:Vintlayers}
&&\mathcal{V}_{\alpha\beta;\gamma\lambda}(\qq;l) \equiv \int \frac{\mathrm{d}k}{2\pi}\, \left(\int\mathrm{d}z\,\phi_{\alpha}(z)\, \phi_{\beta}(z)\,e^{-ikz}\right) \nonumber\\&&\quad\quad\times\,\left(\int\mathrm{d}z' \, \phi_{\gamma}(z'-l) \, \phi_{\lambda}(z'-l) \, e^{ikz} \right) V_{\mathrm{dip}}(k,\qq),\nonumber\\
\end{eqnarray}
where $\phi_\alpha(z)$ is $\alpha$'th harmonic oscillator wavefunction and:
\begin{align}
V_{\mathrm{dip}}(k,\qq) &\equiv \int \mathrm{d}z\,\mathrm{d}^2\xx \, e^{-i k z}\,e^{-i \qq \cdot \xx}\, V_{\mathrm{dip}}(\xx,z)\nonumber\\
&=4 \pi D^2 \left(\frac{k^2}{k^2 + |\qq|^2}-\frac{1}{3}\right),
\end{align}
Note that $\mathcal{V}^{(m,n)}_{\alpha\beta;\gamma\lambda}(\qq) \equiv \mathcal{V}_{\alpha\beta;\gamma\lambda}\left(\qq;(n-m)d\right)$. Evaluating the $k$ integral, we get:
\begin{equation}\label{eq:Vqz}
\int\frac{\mathrm{d}k_z}{2\pi}\,V_{\mathrm{dip}}(k,\qq) = \frac{8\pi D^2}{3}\delta(z-z') - 2 \pi D^2 |\qq| e^{-|\qq||z-z'|}.
\end{equation}
Plugging Eq.~(\ref{eq:Vqz}) into Eq.~(\ref{eq:Vintlayers}), we get:
\begin{multline}\label{eq:Velems}
\mathcal{V}_{\alpha\beta;\gamma\lambda}(\qq;l) = \\
\frac{8\pi D^2}{3}\int \mathrm{d}z\,\phi_{\alpha}(z)\phi_{\beta}(z)\phi_{\gamma}(z-l)\phi_{\lambda}(z-l)\\
- 2 \pi D^2 |\qq| \int \mathrm{d}z\int\mathrm{d}z'\,e^{-|\qq||z-z'|} \phi_{\alpha}(z)\phi_{\beta}(z)\\
\times\,\phi_{\gamma}(z'-l)\phi_{\lambda}(z'-l).
\end{multline}

At this point, one may proceed by finding a generating function for $\mathcal{V}_{\alpha\beta;\gamma\lambda}(\qq;l)$ through expressing the Hermite's functions appearing in the harmonic oscillator wavefunctions in terms of their generating functions~\cite{Babadi2011}. Since we are interested in the first few matrix elements here, we find it is easier to evaluate the required integrals directly. The first integral in Eq.~(\ref{eq:Velems}) is a simple Gaussian integral while the second double integration can be easily evaluated by changing variables to $\eta = (z+z')/2$ and $\xi = (z-z')/2$ and successive integration by parts. We just quote the final result here:
\begin{multline}\label{eq:V0000}
\mathcal{V}_{00;00}(\qq;l) = \frac{4\sqrt{2\pi}D^2}{3\aprp}e^{-l^2/2 \aprp^2} \\
- \pi D^2 |\qq|\,F_+(|\qq|a_\perp,l/a_\perp),
\end{multline}
\begin{multline}\label{eq:V0001}
\mathcal{V}_{00;01}(\qq;l) = -\frac{4\sqrt{\pi}l\,D^2}{3a_\perp^2}\,e^{-l^2/2a_{\perp}^2}\\
+\sqrt{2}\pi D^2 a_\perp |\qq|^2\,F_-(|\qq|a_\perp,l/a_\perp),
\end{multline}
\begin{multline}\label{eq:V0011}
\mathcal{V}_{00;11}(\qq;l) = \frac{2 \sqrt{2\pi} D^2}{3a_\perp}\,e^{-l^2/2a_{\perp}^2}(1+l/a_\perp)\\
+\frac{D^2}{2} |\qq| \Bigg\{2\sqrt{2\pi}|\qq|a_\perp\,e^{-l^2/2a_{\perp}^2} \\
- \pi(2+|\qq|^2 a_\perp^2)\,F_+(|\qq|a_\perp,l/a_\perp),
\end{multline}
\begin{multline}
\mathcal{V}_{01;11}(\qq;l) = \frac{2 \sqrt{\pi} D^2 l}{3a_\perp^2}\,e^{-l^2/2a_{\perp}^2}(l^2/a_\perp^2-1)\\
-\frac{\sqrt{\pi}D^2 |\qq|^2}{4 a_\perp} |\qq| \Bigg\{-4 l/a_\perp\,e^{-l^2/2a_{\perp}^2}\\
+\sqrt{2\pi}(2+|\qq|^2 a_\perp^2)\,F_-(|\qq|a_\perp,l/a_\perp)\Bigg\},
\end{multline}
\begin{multline}\label{eq:V1111}
\mathcal{V}_{11;11}(\qq;l) = \frac{\sqrt{2\pi}D^2}{3\aprp}e^{-l^2/2 \aprp^2}(3-2 l^2/\aprp^2+l^4/\aprp^4) \\
- \frac{D^2}{4}\sqrt{\frac{\pi}{2}} |\qq|\,\Bigg\{-4|\qq|\aprp e^{-l^2/2 \aprp^2}(3+|\qq|^2\aprp^2 + l^2/\aprp^2) \\
+ \sqrt{2\pi}(2+|\qq|^2\aprp^2)^2\,F_+(|\qq|a_\perp,l/a_\perp)\Bigg\}.
\end{multline}
In the above equations, $F_\pm(x,y)$ is defined as:
\begin{multline}
F_\pm(x,y) = e^{-y^2/2}\Bigg[e^{(x-y)^2/2}\,\mathrm{Erfc}\left(\frac{x-y}{\sqrt{2}}\right)\\
\pm e^{(x+y)^2/2}\,\mathrm{Erfc}\left(\frac{x+y}{\sqrt{2}}\right)\Bigg].
\end{multline}

\section{Numerical Solution of the Bethe-Salpeter Equation}\label{app:num}
A major difficulty in evaluating response functions in the TDHF approximation is solving the Bethe-Salpeter equation resulting from the ladder diagram summations, Eq.~(\ref{eq:integeq1}), in order to obtain the irreducible polarization $\Pi^{\star}_{\alpha\beta;\gamma\lambda}$. The bookkeeping of subband indices in quasi-two-dimensional systems is an additional difficulty. Nevertheless, the problem is essentially a system of coupled Fredholm integral equations of the second kind, which can be efficiently solved using numerical methods such as the Nystr\"om method~\cite{Reinhardt1985}. In this method, one approximates the integrations with using quadrature formulas and solves the resulting (large) system of linear equations. 

\begin{figure}[t!]
\includegraphics[width=8cm]{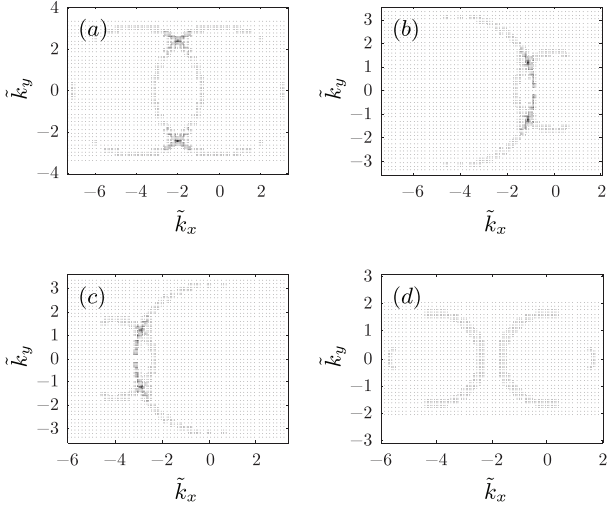}
\caption{Adaptively generated grids for integrations involving (a) $h_{00;\gamma\lambda}$, (b) $h_{01;\gamma\lambda}$, (c) $h_{10;\gamma\lambda}$ and (d) $h_{11;\gamma\lambda}$ for a system in the $\mathrm{N}_1$ phase [$r_d=1.35$, $\sqrt{n}a_\perp=0.25$] and $\sqrt{n}q \approx 4.78$. $\tilde{k}_{x(y)} \equiv n^{-1/2}\,k_{x(y)}$.}
\label{fig:adaptive-grid}
\end{figure}

In the approach used in this study for locating softened modes, one is only interested in the response functions in the static limit, i.e. $\nu_n \rightarrow 0^+$. In this limit, the bare polarizations appearing in Eq.~(\ref{eq:integeq1}) will have integrable singularities at the intersection of particle and hole Fermi surfaces. For example, Fig.~\ref{fig:PIplot} shows the single singularity at $\kk_0 = -k_{F,0}\hat{x}$ for $\qq = 2k_{F,0}\hat{x}$. The dressed polarizations may have additional singularities associated to the softened collective modes. The single-particle and collective singularities can be separated by simply dividing the irreducible polarization by bare polarization. We define:
\begin{multline}
h_{\alpha\beta;\gamma\lambda}(\qq,i\nu_n;\kk) \equiv \\ \left(\Pi^{(0)}\right)^{-1}_{\alpha\beta;\mu\nu}(\qq,i\nu_n;\kk)\,\Pi^{\star}_{\mu\nu;\gamma\lambda}(\qq,i\nu_n;\kk),
\end{multline}
where the inverse of the bare polarization is defined only with respect to subband indices, i.e. $\left(\Pi^{(0)}\right)^{-1}_{\alpha\beta;\mu\nu}(\qq,i\nu_n;\kk)\,\Pi^{(0)}_{\mu\nu;\gamma\lambda}(\qq,i\nu_n;\kk) = \delta_{\alpha\gamma}\delta_{\beta\lambda}$. Recasting Eq.~(\ref{eq:integeq1}) in terms of $h_{\alpha\beta;\gamma\lambda}$, we get:
\begin{multline}\label{eq:integeq2}
h_{\alpha\beta;\gamma\lambda}(\qq,i\nu_n;\kk_1) = \delta_{\alpha\gamma}\delta_{\beta\lambda} -\int\frac{\mathrm{d}^2\kk'}{(2\pi)^2}\,\mathcal{V}_{\nu\beta;\alpha\mu}(\kk'-\kk_1)\\
\times\,\Pi^{(0)}_{\mu\nu;\rho\sigma}(\qq,i\nu_n;\kk')\,h_{\rho\sigma;\gamma\lambda}(\qq,i\nu_n;\kk').
\end{multline}
In the absence of interactions, $h_{\alpha\beta;\gamma\lambda}$ is simply the identity operator in the space of subband indices. Since the bare polarization appears in the integrand, the integrable singularities associated to gapless particle-hole excitations will not result in any singularity in $h_{\alpha\beta;\gamma\lambda}$. On the other hand, if the Fredholm determinant of the above integral equation vanishes at some $\qq$, i.e. $\det\left[\mathbf{1}+\mathcal{V}\,\Pi^{(0)}\right]=0$, $h_{\alpha\beta;\gamma\lambda}$ will be singular at that $\qq$. In fact, this condition can be used as a practical criterion for locating the softened collective modes. Thus, the single-particle poles are absent in $h_{\alpha\beta;\gamma\lambda}$ and it effectively represents the many-body correctiosn to the bare polarization.\\

For non-smooth integral kernels, as it is the case here, rapid convergence of Nystr\"om mehod is only achieved if one employs adaptively generated integration quadratures that properly handle the integrable singularities and fast variations of the integral kernels. The singular points must be avoided and a finer mesh must be used in the proximity of the singularities and sharp variations of the integrand. We implemented the adaptive mesh refinement (AMR) algorithm described in Ref.~\cite{Henk} on a square-based mesh to generate the integration quadrature. For each $\qq$, a uniform rectangular grid was generated and adaptively refined until the relative integration error was smaller than $10^{-4}$. One may generate a single ``global'' quadrature that handles the irregularities of the various integral kernels appearing in Eq.~(\ref{eq:integeq2}) corresponding to different choices of subband indices. However, a more efficient approach can be devised by utilizing the parity conserving nature of intra-layer interactions. For instance, when only the first two subbands are relevant, there is no subband hybridization and $\Pi^{(0)}_{\alpha\beta;\gamma\lambda} \propto \delta_{\alpha\gamma}\delta_{\beta\lambda}\,\Pi^{(0)}_{\alpha\beta}$. Therefore, $h_{\rho\sigma;\gamma\lambda}$ only appears in conjuction with $\Pi^{(0)}_{\rho\sigma}$ in the intergand of Eq.~(\ref{eq:integeq2}) and consequently, one may produce four separate integration quadratures for $h_{00;\gamma\lambda}$, $h_{01;\gamma\lambda}$, $h_{10;\gamma\lambda}$ and $h_{11;\gamma\lambda}$, each of which has about half the number of points of a globally applicable quadrature. Fig.~\ref{fig:adaptive-grid} shows an instance of the adaptive grid generated in this fashion.

In all of the studied cases, the algorithm produced a mesh containing $\sim 5000$ (or less) points before the stopping criteria was fulfilled. The integrals appearing in Eq.~(\ref{eq:integeq2}) was then approximated using the generated quadrature and reduced to a linear system. The linear system was solved using LU decomposition. Once $h_{\alpha\beta;\gamma\lambda}$ was calculated, the irreducible polarization diagrams were finally evaluated by multiplying $h_{\alpha\beta;\gamma\lambda}$ by the bare polarization and summing over $\kk_1$:
\begin{multline}
\Pi^{\star}_{\alpha\beta;\gamma\lambda}(\qq,i\nu_n) = \int \frac{\mathrm{d}^2 \kk_1}{(2\pi)^2}\,\Pi^{(0)}_{\alpha\beta;\mu\nu}(\qq,i\nu_n;\kk_1)\\
\times\,h_{\mu\nu;\gamma\lambda}(\qq,i\nu_n;\kk_1).
\end{multline}  
The previous generated quadratures can be utilized to evaluate the above integral as well.



\begin{thebibliography}{100} 

	\bibitem{Lewenstein2007} M. Lewenstein, A. Sanpera, V. Ahufinger, B. Damski, A. Sen, U. Sen, Adv. Phys. {\bf 56}, 243 (2007).

	\bibitem{Bloch2008} I. Bloch, J. Dalibard, W. Zwerger, Rev. Mod. Phys. {\bf 80}, 885 (2008).

	\bibitem{Ketterle2008} W. Ketterle and M. W. Zwierlein, arXiv:0801.2500v1.

	\bibitem{Lang2008} F. Lang, K. Winkler, C. Strauss, R. Grimm, and J. Hecker Denschlag, Phys. Rev. Lett. {\bf 101}, 133005 (2008).

	\bibitem{Deiglmayr2008} J. Deiglmayr, A. Grochola, M. Repp, K. M\"ortlbauer, C. Gl\"uck, J. Lange, O. Dulieu, R. Wester and M. Weidem\"uller, Phys. Rev. Lett. {\bf 101}, 133004 (2008).

	\bibitem{Ospelkaus2008} S. Ospelkaus, A. Pe�er, K.-K. Ni, J. J. Zirbel, B. Neyen- huis, S. Kotochigova, P. S. Julienne, J. Ye, and D. S. Jin,
Nat. Phys. {\bf 4}, 622 (2008).

	\bibitem{Ni2008} K.-K. Ni, S. Ospelkaus, M. H. G. de Miranda, A. Pe�er,
B. Neyenhuis, J. J. Zirbel, S. Kotochigova, P. S. Julienne,
D. S. Jin, and J. Ye, Science {\bf 322}, 231 (2008).

	\bibitem{Ni2009} K.-K. Ni, S. Ospelkaus, D.J. Nesbitt, J. Ye and D. S. Jin,
Phys. Chem. Chem. Phys {\bf 11}, 9626 (2009).

	\bibitem{Ospelkaus2010} S. Ospelkaus, K.-K. Ni, D. Wang, M. H. G. de Miranda, B. Neyenhuis, G. Qu\'em\'ener, P. S. Julienne, J. L. Bohn, D. S. Jin and J. Ye, Science {\bf 327}, 853-857 (2010).

	\bibitem{Ni2010} K.-K. Ni, S. Ospelkaus, D. Wang, G. Qu\'em\'ener, B. Neyenhuis, M. H. G. de Miranda, J. L. Bohn, J. Ye and D. S. Jin,
Nature {\bf 464}, 1324-1328 (2010).

	\bibitem{Zuchowski2010} P. S. Zuchowski and J. M. Hutson, Phys. Rev. A {\bf 81}, 060703(R) (2010).

	\bibitem{Quemener2010} G. Qu\'em\'ener and J. L. Bohn, Phys. Rev. A {\bf 81}, 022702 (2010).

	\bibitem{Quemener2011} G. Qu\'em\'ener and J. L. Bohn, Phys. Rev. A {\bf 83}, 012705 (2011).

	\bibitem{Yamaguchi2010} Y. Yamaguchi, T. Sogo, T. Ito, and T. Miyakawa, Phys. Rev. A {\bf 82}, 013643 (2010).

	\bibitem{Sun2010} K. Sun, C. Wu, and S. Das Sarma, Phys. Rev. B {\bf 82}, 075105 (2010).

	\bibitem{Baym1961} G. Baym and L. P. Kadanoff, Phys. Rev. {\bf 124}, 287Ð299 (1961).

	\bibitem{NozieresPines} P. Nozieres and D. Pines, {\em Theory of Quantum Liquids}, Westview Press (1999).

	\bibitem{Babadi2011} M. Babadi and E. Demler, arXiv:1106.4345 ({\em submitted to Phys. Rev. A}).

	\bibitem{Overhauser1962} A. W. Overhauser, Phys. Rev. {\bf 128}, 1437Ð1452 (1962).

	\bibitem{Meyer2009} J. S. Meyer and K. A. Matveev, J. Phys.: Condens. Matter {\bf 21}, 023203 (2009).

	\bibitem{Fishman2008} S. Fishman, G. De Chiara, T. Calarco and G. Morigi, Phys. Rev. B {\bf 77}, 064111 (2008).

	\bibitem{Astrakharchik2008} G. E. Astrakharchik, G. Morigi, G. De Chiara and J. Boronat, Phys. Rev. A {\bf 78}, 063622 (2008).

	\bibitem{Shimshoni2011} E. Shimshoni, G. Morigi and S. Fishman, Phys. Rev. A {\bf 83}, 032308 (2011).

	\bibitem{Meng2011} T. Meng, M. Dixit, M. Garst and J. S. Meyer, Phys. Rev. B {\bf 83}, 125323 (2011).

	\bibitem{Potter2010} A. Potter, E. Berg, D.-W. Wang, B. Halperin and E. Demler, Phys. Rev. Lett. {\bf 105}, 220406 (2010).

	\bibitem{Pikovski2010} A. Pikovski, M. Klawunn, G. V. Shlyapnikov and L. Santos, Phys. Rev. Lett. {\bf 105}, 215302 (2010).

	\bibitem{Baranov2011} M. Baranov, A. Micheli, S. Ronen and P. Zoller, Phys. Rev. A {\bf 83}, 043602 (2011).

	\bibitem{Mahan} G. D. Mahan, {\em Many-Particle Physics}, 3rd ed. (Kluwer Academic/Plenum, New York, 1981).

	\bibitem{Yarlagadda1994} S. Yarlagadda and G. F. Giuliani, Phys. Rev. B {\bf 49}, 14188-14196 (1994).

	\bibitem{Trail2003} J. R. Trail, M. D. Towler and R. J. Needs, Phys. Rev. B {\bf 68}, 045107 (2003).

	\bibitem{Tanatar1989} B. Tanatar and D. M. Ceperley, Phys. Rev. B {\bf 39}, 5005 (1989).

	\bibitem{FranzMillis1998} M. Franz and A. J. Millis, Phys. Rev. B {\bf 58}, 14572 (1998).

	\bibitem{BergAltman2007} E. Berg and E. Altman, Phys. Rev. Lett. {\bf 99}, 247001 (2007).

	\bibitem{Spivak2004} B. Spivak and S. A. Kivelson, Phys. Rev. B {\bf 70}, 155114 (2004).

	\bibitem{Lipparini1995} K. Takayanagi and E. Lipparini, Phys. Rev. B {\bf 52}, 1738 (1995).

	\bibitem{Lipparini1996} K. Takayanagi and E. Lipparini, Phys. Rev. B {\bf 54}, 8122 (1996).

	\bibitem{Ito2004} Y. Ito, K. Okazaki and Y. Teraoka, Physica E {\bf 22}, 148Ð151(2004).

	\bibitem{Chaikin2000} P. M. Chaikin and T. C. Lubensky, {\it Principles of Condensed Matter Physics}, Cambridge University Press; Reprint edition (October 9, 2000).

	\bibitem{Sage2005} J. M. Sage, S. Sainis, T. Bergeman and D. DeMille, Phys. Rev. Lett. {\bf 94}, 203001 (2005).

	\bibitem{BCSestimate} Ref.~\cite{Potter2010,Pikovski2010,Baranov2011} have studied the inter-layer superfluidity in the single-subband limit. In order to use their results to estimate the critical superfluid transition temperature for multi-subband gases, the effective inter-layer separation must be reduced according to the number of populated subbands. For instance, in a system with the first two subbands occupied, the minimum distance between the dipoles across two adjacent layers is $d_{\mathrm{eff.}} \approx d - 2a_\perp$.

	\bibitem{Bruun2008} G. M. Bruun and E. Taylor, Phys. Rev. Lett. {\bf 101}, 245301 (2008).

	\bibitem{Zinner2011} N. T. Zinner and G. M. Bruun, arXiv:1102.1551.

	\bibitem{Reinhardt1985} H. J. Reinhardt, {\it Analysis of approximation methods for differential and integral equations}, Springer-Verlag (1985).

\end{thebibliography}
\end{document}